\documentclass{aa}

\usepackage{epsfig}
\usepackage{graphicx} 
\usepackage{float}

\def\gtaprx {\lower .1ex\hbox{\rlap{\raise .6ex\hbox{\hskip .3ex
 {\ifmmode{\scriptscriptstyle >}\else {$\scriptscriptstyle >$}\fi}}}
 \kern -.4ex{\ifmmode{\scriptscriptstyle \sim}\else
 {$\scriptscriptstyle\sim$}\fi}}} 
\def\ltaprx {\lower .1ex\hbox{\rlap{\raise .6ex\hbox{\hskip .3ex
 {\ifmmode{\scriptscriptstyle <}\else {$\scriptscriptstyle <$}\fi}}}
 \kern -.4ex{\ifmmode{\scriptscriptstyle \sim}\else
 {$\scriptscriptstyle\sim$}\fi}}} 
\def\etal {et al. }
\def\littleprime{\ifmmode{\scriptscriptstyle \prime }
 \else{\hbox{$\scriptscriptstyle \prime$ }}\fi}
\def\littless{\ifmmode{\scriptscriptstyle s }
 \else{\hbox{$\scriptscriptstyle s $ }}\fi}
\def\littlemm{\ifmmode{\scriptscriptstyle m }
 \else{\hbox{$\scriptscriptstyle m $ }}\fi}
\def\littlehh{\ifmmode{\scriptscriptstyle h }
 \else{\hbox{$\scriptscriptstyle h $ }}\fi}
\def\littlecirc{\ifmmode{\scriptscriptstyle \circ }
 \else{\hbox{$\scriptscriptstyle \circ $ }}\fi} 
\def\rasec{\raise .9ex \hbox{\littless}} 
\def\arcsec{\raise .9ex \hbox{\littleprime\hskip-3pt\littleprime\hskip-3pt}} 
\def\ramin{\raise .9ex \hbox{\littlemm}} 
\def\arcmin{\raise .9ex \hbox{\littleprime}}
\def\hrs{\raise .9ex \hbox{\littlehh}} 
\def\deg{\hbox{$^\circ$}}
\def\degree{\raise .9ex \hbox{\littlecirc}} 
\def\magpoint{\hbox to 2pt{}\rlap{\hskip -.5ex \arcmm}.\hbox to 2pt{}} 
\def\arcsspoint{\hbox to 1pt{}\rlap{\arcss}.\hbox to 2pt{}} 
\def\arcsecpoint{\hbox to 1pt{}\rlap{\arcsec}.\hbox to 2pt{}} 
\def\arcminpoint{\hbox to 1pt{}\rlap{\arcmin}.\hbox to 2pt{}} 
\def\degreepoint{\hbox to 1pt{}\rlap{\degree}.\hbox to 2pt{}}

\def\lax{{$\mathrel{\hbox{\rlap{\hbox{\lower4pt\hbox{$\sim$}}}\hbox{$<$}}}$}}
\def\gax{{$\mathrel{\hbox{\rlap{\hbox{\lower4pt\hbox{$\sim$}}}\hbox{$>$}}}$}}

\begin{document}

                                                                             

\title{A Radio Census of Nuclear Activity in Nearby Galaxies}

\author{Mercedes  E.  Filho \inst{1,2,3}, Peter  D. Barthel \inst{3} \and
  Luis C. Ho  \inst{4}}

\institute {Centro de Astrof\'\i sica da
  Universidade do Porto, Rua das Estrelas, 4150 -- 762 Porto, Portugal
\and
Istituto di Radioastronomia, CNR, Via P. Gobetti, 101, 40129 Bologna, Italy
\and
Kapteyn  Astronomical  Institute, P.O.~Box 800, 9700
  AV Groningen, The Netherlands
\and
The Observatories of the Carnegie Institution of Washington, 813
  Santa Barbara Street, Pasadena, CA 91101, USA}

\date{Received XX; accepted XX}  


\abstract{In  order  to   determine  the  incidence  of  black  hole
accretion-driven nuclear activity in nearby galaxies, as manifested by
their radio  emission, we  have carried out  a high-resolution Multi-Element 
Radio-Linked Interferometer Network (MERLIN)
survey  of  LINERs  and  composite  LINER/H{\sc ii}  galaxies  from  a
complete magnitude-limited  sample of bright  nearby galaxies (Palomar
sample)  with  unknown arcsecond-scale  radio  properties.  There  are
fifteen radio  detections, of  which three are  new subarcsecond-scale
radio core detections, all being candidate AGN.  The detected galaxies
supplement  the already  known low-luminosity AGN  --  low-luminosity
Seyferts,  LINERs and  composite LINER/H{\sc  ii} galaxies  --  in the
Palomar sample.  Combining all radio-detected Seyferts,  LINERs and  
composite LINER/H{\sc  ii} galaxies 
(LTS sources), we
obtain an  overall radio  detection rate of  54\% (22\% of  all bright
nearby galaxies) and we estimate that at least $\sim$50\% ($\sim$20\% of all bright
nearby galaxies)  are true AGN.   The
radio  powers of the LTS galaxies allow the construction of a local radio
luminosity  function.   
By comparing the luminosity function with those of selected
moderate-redshift AGN,  selected from the 2dF/NVSS survey, we find that 
LTS sources naturally extend the RLF of powerful
AGN down to powers of about 10 times that of Sgr A*.

}                    

\titlerunning {A Radio Census of Nuclear Activity in Nearby Galaxies}

\authorrunning{Filho \etal} 

\maketitle
 

\section{Introduction}

The search for low-luminosity active galactic nuclei (LLAGN) in nearby
galaxies has been  the subject of many optical  surveys.  Results show
that nuclear activity may be  a common phenomenon.  The Palomar survey
(Ho, Filippenko  \& Sargent  1995, 1997a, b)  has been very  useful in
this  regard  by  providing  a  sensitive  magnitude-limited  (B$_{\rm
T}<$12.5~mag)  sample of almost 500 bright nearby  galaxies. 
About  half  of the
sources are  emission-line nuclei,  classified as Seyferts,  LINERs or
composite LINER/H{\sc ii} galaxies,  the last category displaying both
LINER and H{\sc ii}  properties.  However, characterizing the powering
mechanisms  of the  sources  is not  straightforward, particularly  in
low-luminosity sources.  Many of these galaxies possess circum-nuclear
star-forming  regions which  blend with  and  may even  drown out  the
presence of a weak active galactic nucleus (AGN).

Optimally it is necessary to  pick spectral regions where the contrast
between  any hypothetical LLAGN  component and  circum-nuclear stellar
component  is maximized. X-rays are very useful in this regard as shown
by the hard X-ray studies of LLAGN  (Terashima \etal  2000; Terashima,  
Ho \& Ptak 2000; Ho \etal  2001; Terashima \&  Wilson 2003). 
In  the  absence of  spectral  AGN signatures  
such  as  a Seyfert  or
quasar-type continuum or broad  emission lines, radio observations can
offer  an alternative method  for determining  the LLAGN  incidence in
nearby  galaxies.   Measurements  of  radio flux,  compactness,  radio
spectral   index\footnote{F$_{\nu}   \propto  \nu^{-\alpha}$
throughout.}   and  brightness   temperatures  provide  the  necessary
diagnostic tools for determining the nature of the radio emission.



\begin{table*}

\setcounter{table}{0}

\scriptsize

\begin{center}

\begin{minipage}[!ht]{119mm}

\caption{Target MERLIN sources. 
Col.  1  Source name.
Col. 2  and 3: Optical position from  NASA/IPAC Extragalactic Database (NED).  
Col. 4: Adopted distance  from  Ho,  Filippenko \& Sargent (1997a) 
with H$_{\rm 0}$=75 km s$^{-1}$ Mpc$^{-1}$.  
Col. 5: Spectral Type from Ho, Filippenko \& Sargent (1997a). L = LINERs, S =
Seyferts, and T  = composite LINER/H{\sc ii} galaxies.  Colons refer to
uncertain (:)  or highly uncertain (::)  spectral classification.  The
number 1.9 refers  to the presence of broad H$\alpha$ detection and  2 refers to the
absence of  broad H$\alpha$  emission. 
Col. 6:  Hubble type  from Ho, Filippenko \& Sargent (1997a).}

\begin{tabular}[h!]{l c c c c c r}

\hline
\hline

   & RA(J2000)  & Dec(J2000) & $D$   &  &  \\

Galaxy &  (\hrs \ramin \rasec) & (\degree \arcmin \arcsec~) &
(Mpc) & Spectral Type & Hubble Type  \\

 (1)    & (2)     & (3)              & (4)         & (5)   & (6)  \\

\hline

IC\,356   & 04 07 46.9 & $+$69 48 45 & 18.1   & T2   & SA(s)ab pec  \\
IC\,520   & 08 53 42.2 & $+$73 29 27 & 47.0   & T2:    & SAB(rs)ab?  \\
NGC\,428  & 01 12 55.6 & $+$00 58 54 & 14.9 & L2/T2: & SAB(s)m    \\
NGC\,488  & 01 21 46.8 & $+$05 15 25 & 29.3 &  T2:: & SA(r)b   \\
NGC\,521  & 01 24 33.8 & $+$01 43 52 & 67.0 & T2/H: & SB(r)bc       \\
NGC\,718  & 01 53 13.3 & $+$04 11 44 & 21.4 & L2 & SAB(s)a  \\
NGC\,777  & 02 00 14.9 & $+$31 25 46 & 66.5 & S2/L2:: & E1     \\
NGC\,841  & 02 11 17.3 & $+$37 29 50 & 59.5 & L1.9: & (R')SAB(s)ab \\  
NGC\,1169 & 03 03 34.7 & $+$46 23 09 & 33.7   & L2   & SAB(r)b \\
NGC\,1961 & 05 42 04.8 & $+$69 22 43 & 53.1   & L2   & SAB(rs)c\\ 
NGC\,2336 & 07 27 03.7 & $+$80 10 42 & 2.9 & L2/S2 & SAB(r)bc\\
NGC\,2681 & 08 53 32.8 & $+$51 18 50 & 13.3 & L1.9 & (R')SAB(rs)0/a \\
NGC\,2768 & 09 11 37.5 & $+$60 02 15 & 23.7 & L2 & E6: \\
NGC\,2832 & 09 19 46.9 & $+$33 44 59 & 91.6 & L2:: & E$+$2: \\
NGC\,2841 & 09 22 02.6 & $+$50 58 35 & 12.0 & L2 & SA(r)b: \\
NGC\,2985 & 09 50 21.6 & $+$72 16 44 & 22.4 & T1.9 & (R')SA(rs)ab   \\
NGC\,3166 & 10 13 45.6 & $+$03 25 32 & 22.0 & L2 & SAB(rs)0/a     \\
NGC\,3169 & 10 14 15.0 & $+$03 27 57 & 19.7 & L2 & SA(s)a pec    \\
NGC\,3190 & 10 18 05.8 & $+$21 49 56 & 22.4 & L2 & SA(s)a pec spin \\
NGC\,3226 & 10 23 27.0 & $+$19 53 54 & 23.4 & L1.9 & E2: pec    \\
NGC\,3301 & 10 36 55.8 & $+$21 52 55 & 23.3 & L2 & (R')SB(rs)0/a     \\
NGC\,3368 (M\,96) & 10 46 45.7 & $+$11 49 12 & 8.1 & L2 & SAB(rs)ab  \\
NGC\,3414 & 10 51 16.2 & $+$27 58 30 & 24.9 & L2 & S0 pec   \\
NGC\,3433 & 10 52 03.9 & $+$10 08 54 & 39.5 & L2/T2:: & SA(s)c \\
NGC\,3507 & 11 03 25.4 & $+$18 08 12 & 19.8 & L2 & SB(s)b      \\
NGC\,3607 & 11 16 54.3 & $+$18 03 10 & 19.9 & L2 & SA(s)0: \\
NGC\,3623 (M\,65) & 11 18 55.9 & $+$13 05 32 & 7.3 & L2: & SAB(rs)a     \\
NGC\,3626 & 11 20 03.8 & $+$18 21 25 &  26.3 & L2:  & (R)SA(rs)0$+$ \\
NGC\,3627 (M\, 66) & 11 20 15.0 & $+$12 59 30 & 6.6 & T2/S2 & SAB(s)b  \\
NGC\,3628 & 11 20 17.0 & $+$13 35 22 & 7.7 & T2 & SAb pec spin \\
NGC\,3646 & 11 22 14.7 & $+$20 12 31 & 56.8 & H/T2: & SB(s)a  \\
NGC\,3675 & 11 26 07.9 & $+$43 35 10 & 12.8  & T2  & SA(S)b \\
NGC\,3718 & 11 32 35.3 & $+$53 04 01 & 17.0 & L1.9 & SB(s)a pec    \\
NGC\,3780 & 11 39 22.3 & $+$56 16 14 & 37.2 & L2:: & SA(s)c:   \\
NGC\,3884$^a$ & 11 46 12.2 & $+$20 23 30 & 91.6 & L1.9 & SA(r)0/a;   \\
NGC\,3898 & 11 49 15.2 & $+$56 05 04 & 21.9 & T2 & SA(s)ab;  \\
NGC\,3945 & 11 53 13.6 & $+$60 40 32 & 22.5 & L2 & SB(rs)0$+$    \\ 
NGC\,4013 & 11 58 31.3 & $+$43 56 49 & 17.0 & T2 & SAb     \\
NGC\,4036 & 12 01 26.9 & $+$61 53 44 & 24.6 & L1.9 & S0$-$   \\
NGC\,4143 & 12 09 36.1 & $+$42 32 03 & 17.0 & L1.9 & SAB(s)0   \\
NGC\,4203 & 12 15 05.0 & $+$33 11 50 & 9.7 & L1.9 & SAB0$-$:   \\
NGC\,4293 & 12 21 12.8 & $+$18 22 58 & 17.0 & L2 & (R)SB(s)0/a  \\
NGC\,4321 (M\,100) & 12 22 54.9 & $+$15 49 21 & 16.8 & T2 & SAB(s)bc  \\
NGC\,4414 & 12 26 27.1 & $+$31 13 24 & 9.7 & T2: & SA(rs)c?  \\
NGC\,4435 & 12 27 40.5 & $+$13 04 44 & 16.8 & T2/H:  & SB(s)0  \\
NGC\,4438 & 12 27 45.6 & $+$13 00 32 & 16.8 & L1.9 & SA(s)0/a \\
NGC\,4450 & 12 28 29.5 & $+$17 05 06 & 16.8 & L1.9 & SA(s)ab  \\
NGC\,4457 & 12 28 59.1 & $+$03 34 14 & 17.4 & L2 & (R)SAB(s)0/a  \\
NGC\,4548 (M\,91) & 12 35 26.4 & $+$14 29 47 & 16.8 & L2 & SBb(rs)   \\
NGC\,4589 & 12 37 25.0 & $+$74 11 31 & 30.0 & L2 & E2 \\
NGC\,4636 & 12 42 50.0 & $+$02 41 17 & 17.0 & L1.9 & E0$+$    \\
NGC\,4736 (M\,94) & 12 50 53.0 & $+$41 07 14 & 4.3 & L2 & (R)SA(r)ab  \\
NGC\,4750 & 12 50 07.1 & $+$72 52 30 & 26.1 & L1.9 & (R)SA(rs)ab   \\
NGC\,4772 & 12 53 29.0 & $+$02 10 02 & 16.3 & L1.9 & SA(s)a  \\
NGC\,4826 (M\,64) & 12 56 43.7 & $+$21 40 52 & 4.1 & T2 & (R)SA(rs)ab   \\
NGC\,5077$^b$ & 13 19 31.6 & $-$12 39 26 & 40.6 & L1.9 & E3$+$\\
NGC\,5297 & 13 46 23.7 & $+$43 52 20 & 37.8 & L2 & SAB(s)c: spin   \\
NGC\,5322 & 13 49 15.2 & $+$60 11 26 & 31.6 & L2:: & E3$+$ \\
NGC\,5353 & 13 53 26.7 & $+$40 16 59 & 37.8 & L2/T2: & SA0 spin  \\
NGC\,5363 & 13 56 07.1 & $+$05 15 20 & 22.4 & L2 & IA0? \\
NGC\,5371 & 13 55 39.9 & $+$40 27 43 & 37.8 & L2 & SAB(rs)bc \\
NGC\,5377 & 13 56 16.6 & $+$47 14 08 & 31.0 & L2 & (R)SB(s)a    \\
NGC\,5448 & 14 02 49.7 & $+$49 10 21 & 32.6 & L2 & (R)SAB(r)a      \\
NGC\,5678 & 14 32 05.8 & $+$57 55 17 & 35.6 & T2 & SAB(rs)b     \\
NGC\,5813 & 15 01 11.2 & $+$01 42 08 & 28.5 & L2: & E1$+$         \\
NGC\,5838 & 15 05 26.2 & $+$02 05 58 & 28.5 & T2:: & SA0$-$  \\
NGC\,6340 & 17 10 24.9 & $+$72 18 16 & 22.0 & L2 & SA(s)0/a  \\
NGC\,6501 & 17 56 03.7 & $+$18 22 23 & 39.6 & L2:: & SA0$+$:   \\
NGC\,6702 & 18 46 57.6 & $+$45 42 20 & 62.8 & L2:: & E: \\   
NGC\,6703 & 18 47 18.8 & $+$45 33 02 & 35.9 & L2:: & SA0$-$     \\

\hline

\end{tabular}

\smallskip

$^a$ Does not fulfill the magnitude criterion for the Palomar sample.

$^b$ Does not fulfill declination criterion for the Palomar sample.

\end{minipage}

\end{center}

\end{table*}

\normalsize


Several   important  radio   surveys  have   been  conducted   on  the
magnitude-limited Palomar bright  nearby galaxy sample (Ho, Filippenko
\& Sargent 1995, 1997a, b), revealing a large fraction of radio cores,
not only  in ellipticals  but also in  bulge-dominated spirals.   In a
recent  Very Large  Array  (VLA) 5  and 1.4\,GHz,  1\arcsec~resolution
survey of  the low-luminosity  Seyferts of the  Palomar sample  (Ho \&
Ulvestad 2001; HU01 hereafter), it was found  that over 80\% of
the sources harbour a radio core.  In a distance-limited sample of low-
luminosity  Seyferts, LINERs  and composite  LINER/H{\sc  ii} galaxies
observed with  the VLA  at 15\,GHz, 0\arcsecpoint25  resolution (Nagar
\etal 2000, 2002; VLA/N00 and VLA/N02 hereafter; Nagar, Falcke \& Wilson 
2005; VLA/N05 hereafter), it was found that 
$\sim$40\% of the objects harbour
subarcsecond-scale compact radio cores.  A similar study with the VLA,
at  8.4\,GHz, 2\arcsecpoint5~resolution of  all the  composite LINER/H{\sc
ii}  galaxies in  the  Palomar  sample has  been  presented in  Filho,
Barthel \&  Ho (2000, 2002a),  revealing radio cores in  $\sim$25\% of
the  sample sources.   However, although  the radio  core  emission in
these sources  is consistent with the  presence of a  LLAGN, we cannot
exclude  a  stellar origin  from  the  brightness temperature  figures
(T$_{\rm   B}$\lax10$^{5}$~K;   Condon   1992)   obtained   at   these
resolutions.   As  conclusive judgement  requires  Very Long  Baseline
(VLBI)-resolution,  multi-wavelength Very  Long Baseline  Array (VLBA)
and   European  Very  Long   Baseline  Interferometer   Network  (EVN)
observations   have   been  obtained   for   selected  subsamples   of
low-luminosity Seyferts, LINERs and composite LINER/H{\sc ii} galaxies
that showed arcsecond- or subarcsecond-scale radio cores (Falcke \etal
2000; F00 hereafter; Nagar  \etal 2002; VLBA/N02 hereafter; Nagar, Falcke 
\& Wilson 2005; VLBA/N05 hereafter; Ulvestad \&  Ho 2001b; Filho, Barthel  \& Ho
2002b; Filho \etal  2004; Anderson, Ulvestad \& Ho  2004; AU04 hereafter).  In sources
with  subarcsecond-  or  arcsecond-scale  radio  peak  emission  above
2.5~mJy,  results reveal  a 100\%  detection rate  of  high-brightness
temperature   (T$_{\rm   B}$\gax10$^8$~K),   compact,  flat   spectrum
($\alpha$$<$0.5) radio cores, enforcing the LLAGN scenario for the radio
emission (VLBA/N05;
see also Ulvestad \& Ho 2001b; 
Filho, Barthel \& Ho 2002b; Filho \etal 2004; AU04).  
Their low  radio luminosities suggest we are  probing the very
faint end of the AGN population.

Unambiguously  determining the  physical nature  of the  nearby galaxy
radio cores is  more than of mere phenomenological  interest.  If they
truly contain  an accretion-powered nucleus, then  they obviously need
to  be included  in  the AGN  population.   Their non-trivial  numbers
impact on several astrophysical problems ranging from the cosmological
evolution of the AGN luminosity function  to their contribution to the
X-ray background.

The present paper  deals with high-resolution radio-imaging of LLAGN,
carried   out  with  the   Jodrell  Bank   Multi-Element  Radio-Linked
Interferometer  Network  (MERLIN), completing  the  radio census of  nuclear
activity  in the Palomar galaxy survey of  Ho, Filippenko  \& Sargent
(1997a).

\section{Source Selection}

The target sources were taken from the Palomar survey of bright nearby
galaxies  (Ho,  Filippenko  \&  Sargent 1997a).   The  Palomar  sample
contains a total of 417 emission-line objects (486 galaxies in total),
of  which  206  are  stellar-powered  H{\sc  ii}  nuclei.   Among  the
remaining  sources,  52   are  classified  as  low-luminosity  Seyfert
galaxies, 94 as  LINERs and 65 as composite  LINER/H{\sc ii} galaxies,
many of which may be AGN-powered.  After  the
extensive   radio   studies  described   above   which  targeted   all
low-luminosity  Seyfert   and  part   of  the  LINERs   and  composite
LINER/H{\sc ii} galaxies  in the Palomar sample, we  have selected the
remaining (68) LINERs  or composite sources (plus one  Seyfert and one
H{\sc ii} nucleus; see Table~1) having either unknown radio properties
or  known  radio emission  in  excess  of  2~mJy on  arcsecond-scales.
Aiming  to detect (weak)  radio emission  on subarcsecond  scales, the
sources  were  observed  with  the  Jodrell Bank  MERLIN,  at  5\,GHz,
0\arcsecpoint1~resolution.  These MERLIN observations were intended as
a filter for follow-up, high-resolution EVN observations, which should
increase   the  brightness  temperature   figures  to   the  necessary
$>$10$^{5}$~K (Condon 1992)  level and thereby unambiguously determine
the  nature  of the  radio  emission.   In  Table~1 we  summarize  the
properties of the 70 sources observed with MERLIN.

\section {Observations and Data Reduction}

The  5\,GHz  MERLIN observations  were  obtained  during 2001  October
18--20,  with  a  15\,MHz  bandwidth.   Defford,  Cambridge,  Knockin,
Darnhall  and Tabley telescopes  were used,  yielding a  resolution of
about 0\arcsecpoint1 at 5\,GHz.  Typically  four 8 minute scans of the
sources were interspersed with 2  minute scans of the respective phase
calibrators, for a total integration  time of about 25 minutes on each
target  source.  3C\,286,  the primary  flux calibrator,  was observed
twice during the observing run.   Initial calibration, reduction and imaging 
of the MERLIN data
was performed by S. Garrington and A. Richards using the AIPS pipeline
at Jodrell Bank.



\begin{table}[!ht]

\setcounter{table}{1}

\footnotesize

\begin{center}

\begin{minipage}{87mm}

\caption{Map parameters of the MERLIN-detected sources.  
Col. 1: Source name.   
Col. 2: Spectral  class from Ho, Filippenko  \& Sargent (1997a).    
Col. 3: Restoring beam.  
Col. 4: Position angle of the beam.  
Col. 5: $rms$ noise level of the image.}

\begin{tabular}{l c c c c}

\hline
\hline

             & Spectral  & Beamsize  & P.A.  &  $rms$   \\
Galaxy       & Class     & (mas$^2$) & (deg) & (mJy/beam)   \\

(1)          & (2)       & (3)       & (4)   & (5) \\

\hline

N\,2768 & L2 & 130 $\times$ 68 & $-$\,7.07  & 0.10 \\

N\,3169 & L2 & 178 $\times$ 56 & $-$45.27 & 0.09 \\

N\,3226 & L1.9 & 153 $\times$ 48 & $-$51.23 & 0.07 \\

N\,3414 & L2 & 153 $\times$ 49 & $-$49.53 & 0.06 \\ 

N\,3718 & L1.9 & 161 $\times$ 130 & $+$54.66 & 0.06 \\

N\,3884$^{a}$ & L1.9 & 86 $\times$ 58 & $-$79.10 & 0.05 \\

N\,4143 & L1.9 & 71 $\times$ 41 & $+$50.37 & 0.06 \\

N\,4203 & L1.9 & 126 $\times$ 47 & $-$28.85 & 0.06 \\

N\,4293 & L2 & 159 $\times$ 55 & $-$18.22 & 0.06 \\

N\,4321 & T2 & 174 $\times$ 55 & $-$15.41 & 0.06 \\

N\,4450 & L1.9 & 177 $\times$ 57 & $-$14.73 & 0.06 \\


N\,5077$^{a}$ & L1.9 & 145 $\times$ 47 & $+$12.02 & 0.08  \\
 
N\,5297$^{a}$ & L2 & 71 $\times$ 42 & $+$29.53 & 0.06 \\

N\,5353 & L2 & 73 $\times$ 44 & $+$23.66 & 0.06  \\

N\,5363 & L2/T2: & 100 $\times$ 46 & $+$24.83 & 0.11 \\

\hline

\end{tabular}

\smallskip

\scriptsize

$^a$ New subarcsecond-scale radio detection.


\end{minipage}

\end{center}

\end{table}

\normalsize



\setcounter{table}{2}

\begin{table*}[!t]

\footnotesize

\begin{center}

\begin{minipage}{110mm}

\caption{The 5\,GHz radio parameters of the MERLIN-detected sources.
Col. 1: Source name.
Col. 2: 5\,GHz peak radio flux density.
Col. 3 and 4: 5\,GHz radio position.
Col. 5: Integrated 5\,GHz flux density.
Col. 6: Brightness temperature. }

\begin{tabular} {l c c c c c}

\hline
\hline

       & F$_{\rm peak}$ & RA(J2000) & Dec(J2000) &  F$_{\rm int}$  & T$_{\rm B}$ \\

Galaxy & (mJy) & (\hrs \ramin \rasec) &
(\degree \arcmin \arcsec~) & (mJy) & ($\times$ 10$^4$ K) \\

 (1)    & (2)  & (3)       & (4)     & (5) & (6) \\

\hline

NGC\,2768 & 7.06 & 09 11 37.414 & $+$60 02 14.86 & 7.22 & 3.9 \\

NGC\,3169 & 2.23 & 10 14 15.014 &  $+$03 27 57.26 & 2.96 & 1.1 \\

NGC\,3226 & 3.50 & 10 23 27.008 &  $+$19 53 54.67 & 3.88 &  2.3 \\

NGC\,3414 & 1.05 & 10 51 16.208 &  $+$27 58 30.27 & 1.13 &  0.7 \\
 
NGC\,3718 & 1.65 & 11 32 34.854 &  $+$53 04 04.47 & 2.07 &  0.4 \\

NGC\,3884 & 1.74 & 11 46 12.183 &  $+$20 23 29.93 & 1.75 &  1.7 \\

NGC\,4143 & 0.97 & 12 09 36.065 &   $+$42 32 03.00 & 1.00 & 1.6 \\

NGC\,4203 & 4.30 & 12 15 05.052 &  $+$33 11 50.35 & 4.86 &  3.6 \\

NGC\,4293 & 0.54 & 12 21 12.797 &  $+$18 22 57.41 & 1.81 &  0.3 \\

NGC\,4321 & 0.50 & 12 22 54.938 & $+$15 49 20.87 & 0.52 &  0.3 \\

NGC\,4450 & 1.21 &  12 28 29.590 &  $+$17 05 06.00 & 1.23 &  0.6 \\





NGC\,5077A & 65.86 & 13 19 31.726 & $-$12 39 27.88 & 67.37 &  47.3 \\

NGC\,5077B & 85.03 & 13 19 31.603 & $-$12 39 29.11 & 86.66 & 61.1 \\

NGC\,5297 & 0.91 & 13 46 23.772 &   $+$43 52 18.88 & 0.95 & 1.5 \\

NGC\,5353 & 16.21 & 13 53 26.695 &   $+$40 16 58.89 & 17.42 &  24.7 \\

NGC\,5363A & 10.93 &  13 56 07.215 &   $+$05 15 17.15 & 11.92 &  11.6 \\

NGC\,5363B & 10.95 & 13 56 07.213 &   $+$05 15 17.26 & 11.28 &  11.7 \\

\hline 

\end{tabular}

\smallskip

\end{minipage}

\end{center}

\end{table*}

\normalsize


\section{Radio Properties}

Of the 70 objects observed  with MERLIN, fifteen objects were detected
above the  5$\sigma$ ($\sim$0.5~mJy beam$^{-1}$)  threshold. Three of these
are new  radio core  detections on subarcsecond scales. With the
exception  of  the  composite  galaxy  NGC\,4321,  all  of  the detected sources  are
spectroscopically classified as LINERs.    All detected
sources except  for NGC\,4293  and NGC\,4321 show  compact, unresolved
cores near the phase center, coincident with the optical nucleus.

In Table~2  we summarize the radio parameters  for the MERLIN-detected
sources. The  $rms$ noise level  was measured in a  source-free region
using the AIPS task IMSTAT.  MERLIN radio maps of the detected sources
(excluding NGC\,4321)  are shown in Fig.~1--4.  The  map for NGC\,4321
can be  seen in Fig.~5, along with the detected background sources. 
Contour  levels are $rms  \times$($-$3, 3, 6,
12, 24, 48, 96, 192).

Table~3 includes the model fits  to the detected radio components. The
AIPS  task IMFIT  was  used  to fit  bi-dimensional  Gaussians to  the
brightness peaks in each  radio component.  The  brightness temperature 
has been calculated using the formula (Weedman 1986):

\begin{center}
T$_{\rm    B}$    =    4.9$\times$10$^7\left(\frac{\rm    F_{\nu}}{\rm
mJy}\right)\left(\frac{\theta_{\rm major} \, {\rm x} \, \theta_{\rm minor}}{\rm
mas^2}\right)^{-1}\left(\frac{\nu}{\rm 5\,GHz}\right)^{-2}$,
\end{center}

\noindent where F$_{\nu}$ is the  peak flux density of the sources and
$\theta_{\rm major}$ and  $\theta_{\rm minor}$ refer to the major and minor 
axis of the Gaussian  beam (Table 2).

Below we give  a source description, quoting values  from surveys such
as the NVSS,
FIRST, VLA/N05 and VLBA/N05 (see Table~4 for details). 
Sources are said to be 'compact' if the FIRST flux density
is $>$50\% the NVSS value, otherwise the source is said to be resolved. Furthermore,
we assume that any radio variability is small compared to the total NVSS flux.
We  also present a  brief discussion on  detected background sources (Fig.~5).

\smallskip

{\it  NGC\,2768:} This  source  has  a  NVSS  (14.5~mJy)  and a  FIRST
(12.3~mJy)  radio detection, denoting  that the  source is  compact on
arcsecond  scales. NGC\,2768 also has an EVN (7~mJy; P. Barthel, private 
communication 2005) and a VLBA/N05 detection of 7.3 mJy 
Comparison of our MERLIN measurement with that of VLA/N05
yields an $\alpha\sim$0.0.

{\it NGC\,3169:} There is a NVSS (87.0~mJy) and FIRST (16.4~mJy) radio
detection;  the flux  density values  show resolution  effects.  VLA/N00
(see also VLA/05)
has detected a 6.8~mJy  source  and FOO (see also VLBA/N05) detect 6.2~mJy; the source  
is compact on
milliarcsecond scales.  Using the VLA/N00 and F00 (see also VLA/N05 and VLBA/N05) values implies a 
flat  spectral index for this  radio source
($\alpha  \sim$0.0).  However,  comparison  of these  values with  our
MERLIN measurement suggests the source could be radio variable.

{\it NGC\,3226:}  There is no NVSS value for this source, most likely because it is
spatially confused with NGC\,3227. However, there is a
4.5~mJy FIRST  detection. VLA/N00 (see also VLA/N05) and F00 (see also
VLBA/N05) 
detect  a 5.4  and  3.5~mJy radio  source,  respectively, showing  the
source to be compact on  milliarcsecond scales. From these last values
the source appears to have a flat/inverted spectrum ($\alpha \gtaprx-$0.4).
The  F00 measurement  is consistent  with  our MERLIN
value; the source does not appear variable at 5\,GHz.



\begin{table}[!t]

\setcounter{table}{3}

\footnotesize

\begin{center}

\begin{minipage}{70mm}

\caption{Bibliographical radio data for the MERLIN-detected sources. 
Col. 1: Survey.
Col. 2: Observing frequency.
Col. 3: Resolution.   
Col. 4: Instrument.  
Col. 5: Reference.  
}

\begin{tabular}{l c c c c}

\hline
\hline

             & $\nu$     & Res.     &        &     \\
Survey       & (GHz)     & (mas) & Instr. & Ref. \\
(1)          & (2)       & (3)       & (4)   & (5) \\

\hline

NVSS & 1.4 & 45000 & VLA & 1 \\
\hline

FIRST & 1.4 & 5000 & VLA & 2 \\
\hline

VLA/N00 & 15 & 150 & VLA & 3 \\
VLA/N02 & 15 & 150 & VLA    & 4 \\
VLA/N05 & 15 & 150 & VLA     & 5 \\
\hline

VLBA/N02 & 5 & 2 & VLBA & 4 \\
VLBA/N05 & 5  &  2     &  VLBA    & 5 \\
F00      & 5  &  2     &  VLBA    & 6 \\
\hline

HU01 & 5.0 & 1000 & VLA & 7 \\
\hline

AUH04 & 1.7 & 10 & VLBA & 8 \\
      & 2.3 & 6 &      &                               \\       
      & 5.0 & 2 &      &                               \\
      & 8.4 & 1.6 &     &                                \\
      & 15.4 & 1.2 &    &                                \\
      & 22.2 & 0.8 &    &                                 \\
      & 43.2 & 0.4 &    &                                \\
                      
\hline

\end{tabular}

\smallskip

\scriptsize

{\sc References} -- (1) Condon \etal 1998; (2) Becker, White, \& Helfand 1995;
(3) Nagar \etal 2000; (4) Nagar \etal 2002; (5) Nagar, Falcke \& Wilson 2005 and
references therein;
(6) Falcke \etal 2000; (7) Ho \& Ulvestad 2001; (8) Anderson, Ulvestad \& Ho 2004.

\end{minipage}

\end{center}

\end{table}

\normalsize


{\it NGC\,3414:} The NVSS (4.4~mJy) and FIRST (4.0~mJy) flux densities
show that  this source  is compact on  arcsecond scales. Our MERLIN combined
with VLA/N05 (2.4~mJy) yields an $\alpha\sim$0.0.

{\it NGC\,3718:} There is a NVSS (14.9~mJy) and FIRST (10.6~mJy) radio
detection, whose flux density values  show the source to be compact on
arcsecond scales. There  is also a VLA/N02 and VLBA/N02 (10.8 and
5.3~mJy detection, respectively (see also VLA/N05 and VLBA/N05).  Comparison of  
these last values  with our
MERLIN measurement  shows that the  source may suffer  from resolution
effects and/or can be radio variable ($\alpha \gtaprx-$0.6).

{\it  NGC\,3884:} This  source  has  a  NVSS  (14.0~mJy)  and a  FIRST
(8.9~mJy)  radio detection,  showing some  resolution  effects. Taking
into consideration  the resolution  mismatch, comparison of  the FIRST
and our MERLIN detection shows  that the source may be resolved and/or
radio  variable  ($\alpha<$1.3).   Because  it does  not  fulfill  the
magnitude criterion for the Palomar survey, this source is not included
in the final sample.

{\it NGC\,4143:} There  is a NVSS (9.9~mJy) and  FIRST (5.0~mJy) radio
detection,  showing some  signs of  resolution. This  source  was also
observed in VLBA/N02 (8.7~mJy; see also VLBA/N05) 
and twice in the
VLA/N00 and VLA/N02 (3.3~mJy and 10.0~mJy; see also VLA/N05).  Comparison of
these values with our MERLIN measurement shows that the source must be
radio variable.

{\it NGC\,4203:}  This source has  been extensively observed  at radio
frequencies.  There is a  NVSS (6.1~mJy), FIRST (6.9~mJy), VLA/N00 and VLA/N02
(9.5 and 9.0~mJy; see also VLA/N05), F00 (8.9~mJy; see also VLBA/N05), 
HU01 (11.2~mJy) and multifrequency AUH04 
radio detections.  The  flux density values,
including  the MERLIN  detection, show  that the  source  is 
compact and  has a flat spectrum  ($\alpha \sim$0.1).


{\it  NGC\,4293:} This  source  has  a  NVSS  (18.5~mJy)  and a  FIRST
(13.9~mJy) radio  detection, denoting that it is  compact on arcsecond
scales.    The  present   MERLIN  and   VLA/N02 (1.4~mJy; see also VLA/N05)
measurements  show that  most  of  the source  has  been resolved  and
possesses a  flat spectrum ($\alpha \ltaprx$0.2).   The weak amorphous
nature  of the  MERLIN emission,  associated with  the  flat spectrum,
suggests the radio emission to be mainly thermal.

{\it   NGC\,4321,   M\,100:}  There   are   several  radio   frequency
measurements for this source (see source discussion in Filho, Barthel \& 
Ho 2000). NVSS has
detected 85.9~mJy and FIRST a 37.4 mJy source.  The position  of the
MERLIN  radio source  is coincident  with the  optical nucleus  of the
galaxy as given by NED and with component A3 in Collison \etal (1994),
the strongest  component in their  multiple radio structure  map.  The
radio  measurements  clearly  show   the  source  has  been  resolved.
Comparison  of  the  Collison  \etal  (1994)  values  and  our  MERLIN
detection, shows that the source is not variable at 5\,GHz and appears
to have a fairly steep radio spectrum ($\alpha <$ 1.6).

{\it NGC\,4450:} Comparison of the NVSS (9.4~mJy), FIRST (6.7~mJy), HU01  
(6.5~mJy), VLA/N02 (2.7~mJy; see also VLA/N05), multifrequency AUH04
and MERLIN measurements show that  the source is compact on  arcsecond scales but
has  been  partially  resolved  on  subarcsecond  scales.  The AUH04
multifrequency  data  yield  a  spectral  index  of  $\alpha
\ltaprx$0.3.


{\it NGC\,4589:} NVSS has detected a 36.8~mJy radio source, VLA/N05 
data yield 11.9 mJy and VLBA/N05 data yield 11.5 mJy; 
the source is compact on subparsec scales. 
Our MERLIN data suffer from severe phase variations.

{\it NGC\,5077:} The  NVSS has detected a 156.7~mJy  radio source. Our
MERLIN  map shows  two very  strong  radio sources  (A and  B) in  the
nuclear region  of this galaxy.  The  NVSS flux density  appears to be
the sum of  the MERLIN flux of these two  sources, denoting that there
is little  extended flux on arcsecond  scales.  While the  APM and NED
identify  these radio  sources with  the nucleus  of NGC\,5077,  it is
difficult  to  constrain the  origin  of  the  radio emission  without
spectral information. Because it does not fulfill the declination criterion
for the Palomar survey, this source is not included in the final sample.



\begin{figure*}
\leavevmode
\centerline{
\epsfxsize=7.0cm
\epsffile{fig1a_4510.ps}
a)
\epsfxsize=7.0cm
\epsffile{fig1b_4510.ps}
b)
}
\end{figure*}

\begin{figure*}
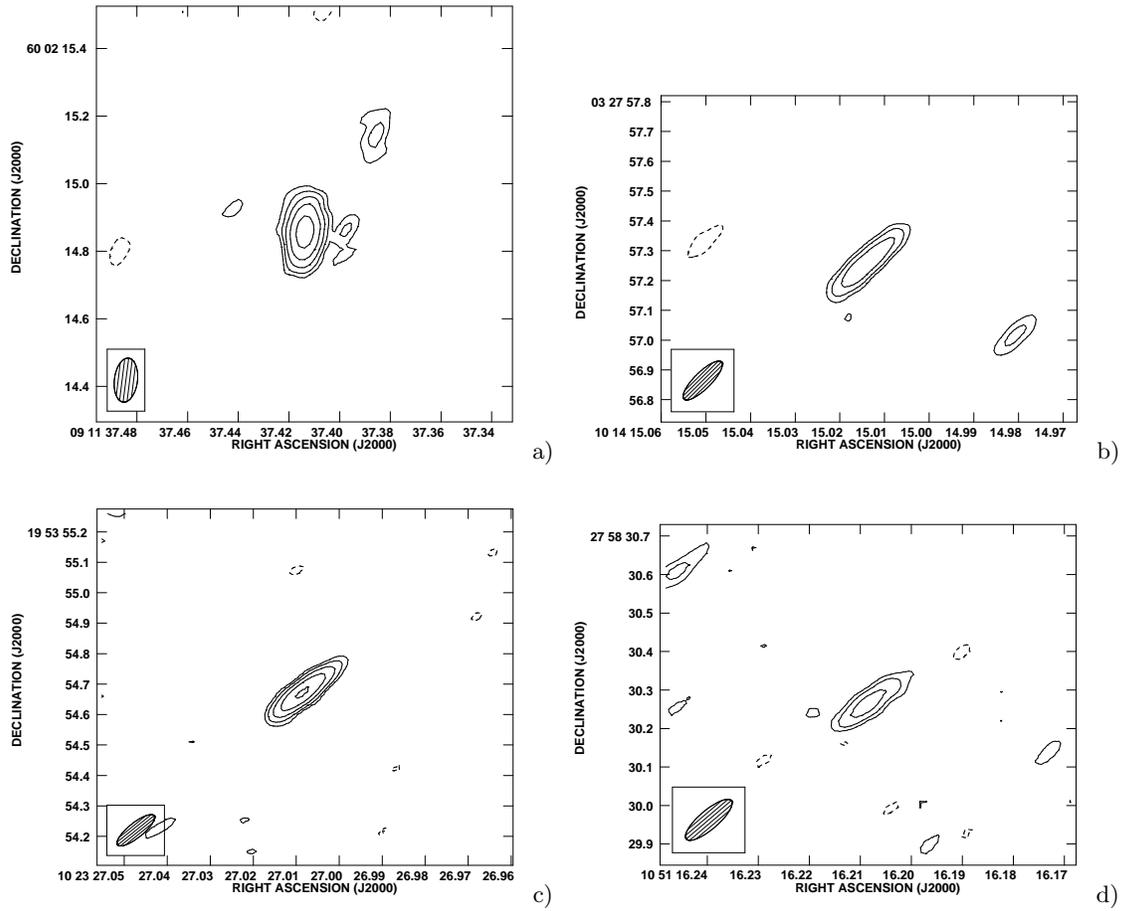

\leavevmode 
\centerline{ 
\epsfxsize=7.0cm  
\epsffile{fig1c_4510.ps} 
c)
\epsfxsize=7.0cm 
\epsffile{fig1d_4510.ps} 
d) 
}
\caption{Radio    emission    contours    of    the    MERLIN-detected
sources. Contour levels are $rms \times$  ($-$3, 3, 6, 12, 24, 48, 96,
192), (see Table~2).  The size of  the restoring beam is given in each
image  at  the  bottom  lefthand  corner  (see  Table~2). {\bf a)}
NGC\,2768,  {\bf b)} NGC\,3169,  {\bf c)}  NGC\,3226  and  {\bf d)}
NGC\,3414.}
\end{figure*}

\clearpage



\begin{figure*}
\leavevmode                
\centerline{               
\epsfxsize=7.0cm
\epsffile{fig2a_4510.ps}            
a)
\epsfxsize=7.0cm
\epsffile{fig2b_4510.ps} 
b)
}
\end{figure*}

\begin{figure*}
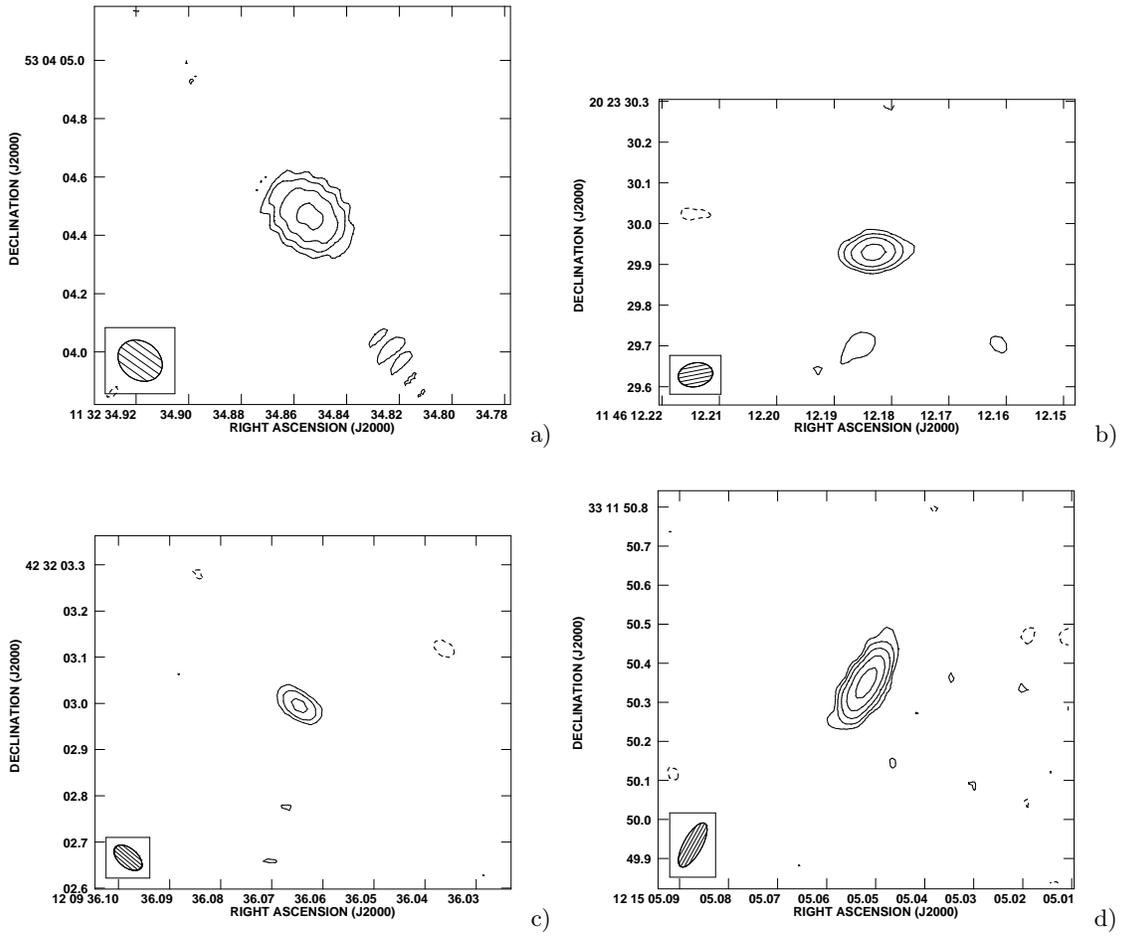

\leavevmode
\centerline{
\epsfxsize=7.0cm
\epsffile{fig2c_4510.ps}
c)
\epsfxsize=7.0cm 
\epsffile{fig2d_4510.ps}
d) 
}
\caption{As  in   Figure~1. {\bf a)} NGC\,3718, {\bf b)}  NGC\,3884, {\bf c)}
NGC\,4143 and {\bf d)} NGC\,4203.}
\end{figure*}

\clearpage



\begin{figure*}
\leavevmode
\centerline{
\epsfxsize=7.0cm
\epsffile{fig3a_4510.ps}
a)
\epsfxsize=7.0cm 
\epsffile{fig3b_4510.ps}
b) 
}
\end{figure*}

\begin{figure*}
\leavevmode
\centerline{
\epsfxsize=7.0cm
\epsffile{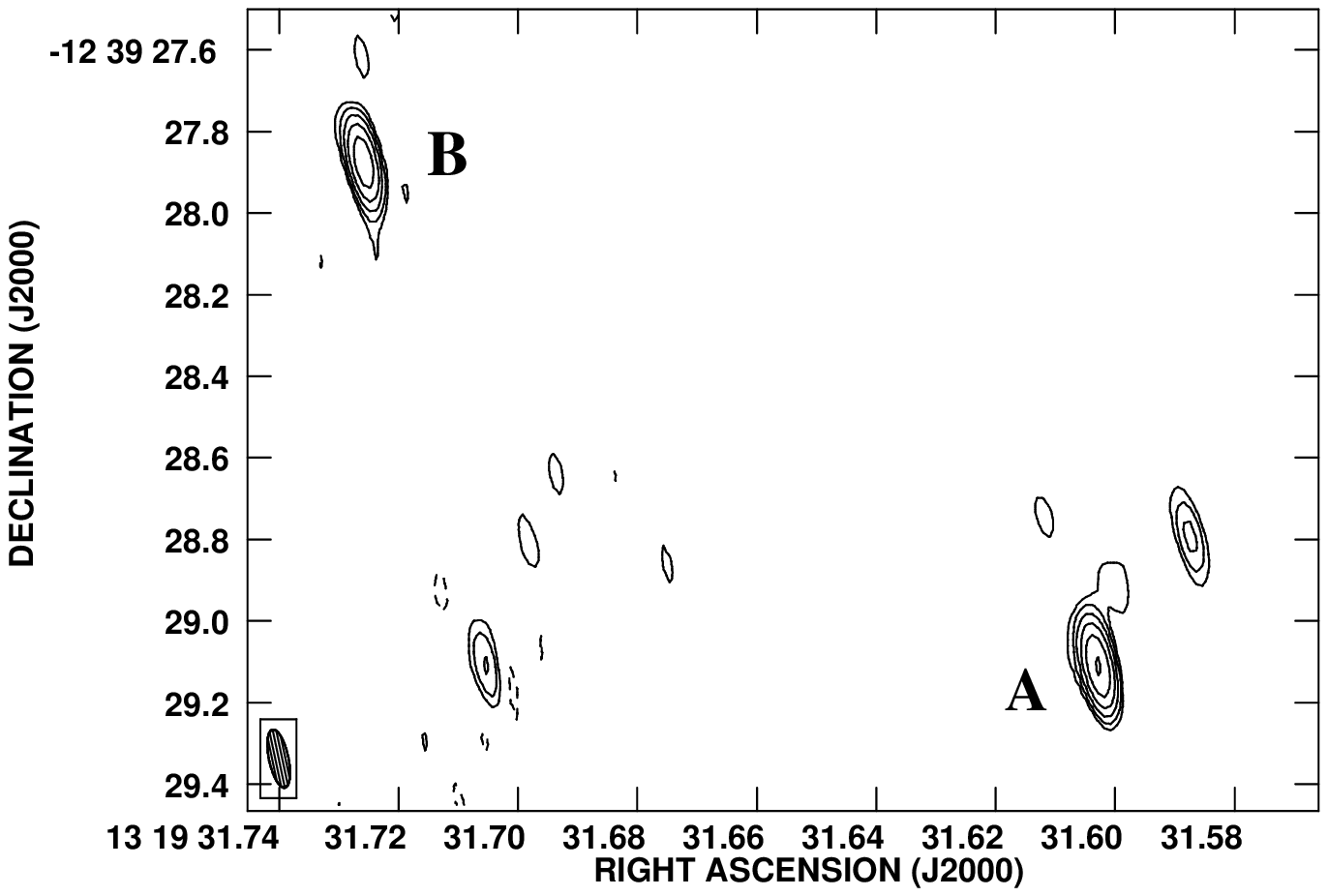}
c)
\epsfxsize=7.0cm
\epsffile{fig3d_4510.ps}
d) 
}
\caption{As  in   Figure~1. {\bf a)} NGC\,4293, {\bf b)} NGC\,4450,
{\bf c)} NGC\,5077 and {\bf d)} NGC\,5297.}
\end{figure*}

\clearpage



\begin{figure*}
\leavevmode
\centerline{
\epsfxsize=7.0cm
\epsffile{fig4a_4510.ps}
a) 
}
\end{figure*}

\begin{figure*}
\leavevmode
\centerline{
\epsfxsize=7.0cm
\epsffile{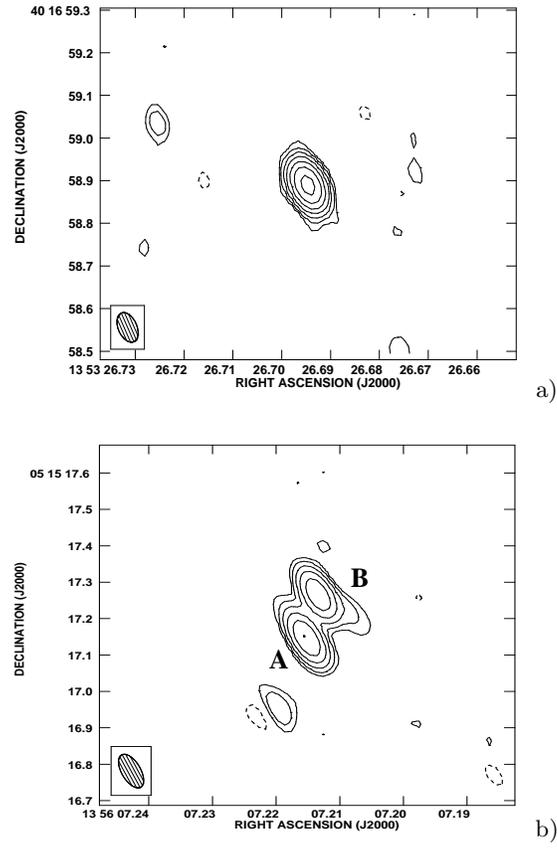}
b)
}
\caption{As  in   Figure~1. {\bf a)} NGC\,5353 and {\bf b)}  NGC\,5363.}

\end{figure*}

\clearpage



\begin{figure*}
\leavevmode
\centerline{
\epsfxsize=7.0cm
\epsffile{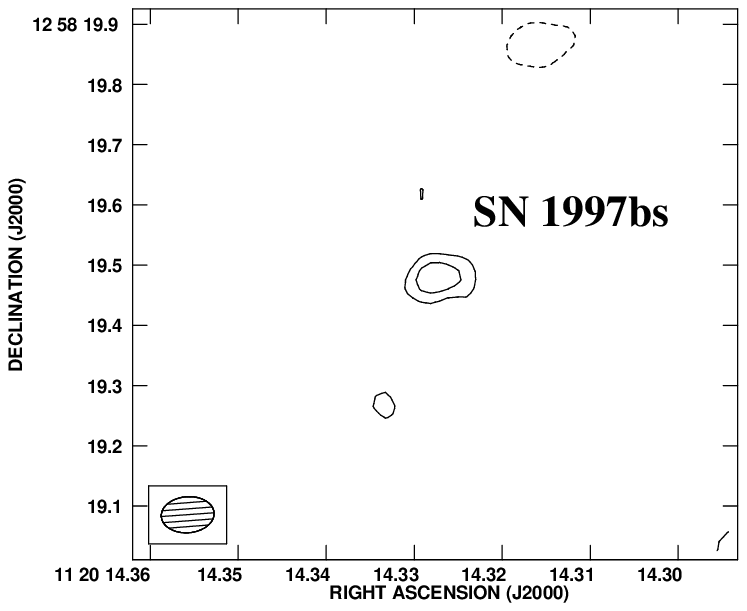}
a)
\epsfxsize=7.0cm
\epsffile{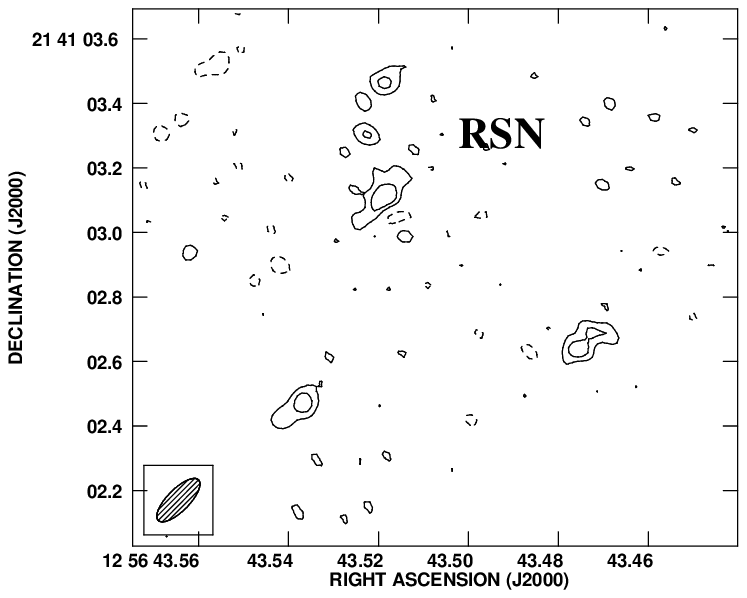}
b)
}
\end{figure*}

\begin{figure*}
\leavevmode  
\centerline{ 
\epsfxsize=5.0cm  
\epsffile{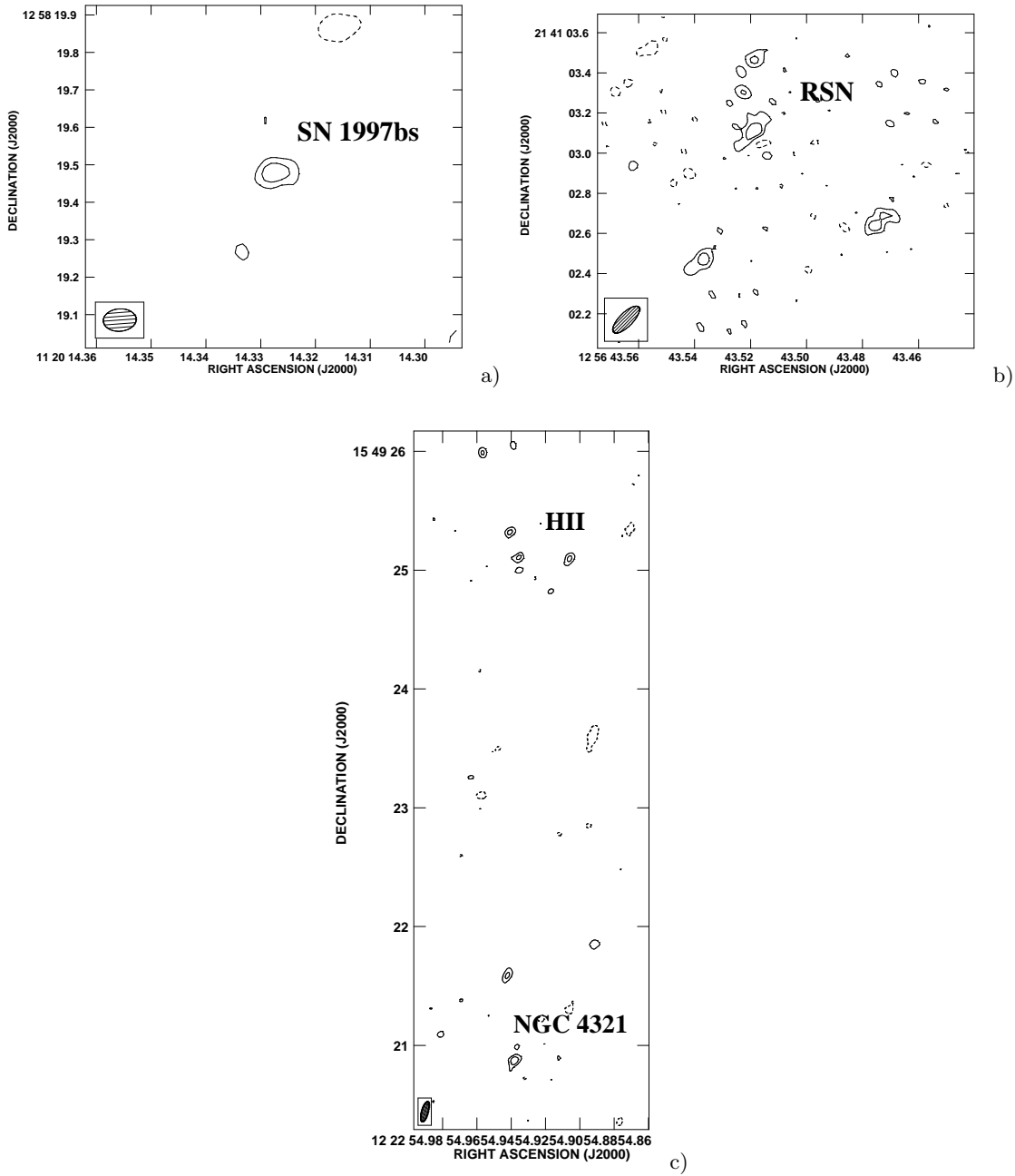} 
c)
}
\caption{MERLIN background  detections (plus NGC\,4321).   Contours as
in Fig.~1. {\bf a)} Source  positionally coincident with SN 1997bs in
NGC\,3627, {\bf b)} source  positionally  coincident  with a  radio
supernova (RSN) in NGC\,4826 (source  number 6; Turner \& Ho 1994) and
{\bf c)} source  positionally coincident  with an  H{\sc  ii} region
(H{\sc ii} region  number 158; Hodge \& Kennicutt  1983) in NGC\,4321,
whose central radio emission can be seen to the South.}
\end{figure*}


{\it NGC\,5297:} NVSS has detected  a 23.0~mJy radio source.  FIRST has
failed to detect this source most likely due to resolution. The  MERLIN 
observations  have resolved out a large part of the radio emission.

{\it  NGC\,5353:} This  source  has  a  NVSS  (39.0~mJy) and FIRST
(38.3~mJy) radio detection, denoting  that the  source is  compact on
arcsecond  scales. There is also a VLBA/N05 (21.6 mJy) and VLA/N05 (18.7~mJy) 
detection.  The MERLIN and VLA/N05 measurements
yield an $\alpha\sim$0.0.

{\it  NGC\,5363:}  The FIRST  (141.8~mJy)  and  NVSS (160.3~mJy)  flux
densities show  that this source  is compact on arcsecond  scales. There is 
also a VLA/N05 and VLBA/N05 measurement of 40.7 and 
39.6 mJy, respectively. The
MERLIN observations have, however, resolved part of the radio emission
into  two components  -- A  and  B, with a total flux density of 
$\sim$25 mJy.  Consultation  of NED  and the  APM
provides  no  other optical  identification  for  these sources  except
sources belonging to the nuclear region of NGC\,5363. It is possible that this 
source has a double nucleus.

\smallskip

{\it Background Sources}

\smallskip

{\it  NGC\,3627:} As  given by  NED,  we may have  detected the  supernova
SN\,1997bs  at  ($\alpha,\delta$)$_{\rm J2000}$ = (11 20 14.2, 
$+$12 58 20) within less than 20\arcsec~of the radio position 
measured in the map of this galaxy.

{\it     NGC\,4321:}    To    the     North,    at  the  NED   position
($\alpha,\delta$)$_{\rm J2000}$ = (12 22 54.9, $+$15 49 25),   
there  is emission  coincident  within less than 20\arcsec~with  the  
H{\sc  ii} region  (source  158),  as documented in Hodge \& Kennicutt (1983).

{\it NGC\,4826:}  We do  not detect the  multiple source  structure as
found by  Turner \&  Ho (1994) because  the individual  sources either
fall below our  detection limit or fall outside  the mapped region. We
do, however,   detect a radio  source  at  the NED position 
($\alpha,\delta$)$_{\rm J2000}$ = (12 56  43.66, $+$21  41 
03),  which  we
identify, within less than 20\arcsec,  with a  RSN (source 6) in 
Turner \&  Ho (1994). Comparison of
our and the Turner \& Ho (1994) fluxes suggest some variability.

\smallskip

Given  the detection  limit  of the  survey  -- 5$\sigma  \sim$0.5~mJy
beam$^{-1}$ -- the overall  detection rate for the MERLIN observations
is 27\%. We have excluded from this calculation the two sources which
do not fulfill the magnitude/declination criteria, the H{\sc ii} and 
Seyfert galaxies (Table~1),
eight sources with incorrect pointing relative to the NED/radio position and
two sources with severe phase errors (see below). We can also
estimate the fraction of MERLIN-detected sources -- 58\% -- that are genuine AGN, due
to the presence of broad H$\alpha$ emission (Type 1.9 sources; see also Section~5). Only
19\% of the MERLIN-detected sources are Type 2, which exhibit
no broad emission lines.  It is  possible  that Type~2
sources are a  mixed case, whereby only a small  fraction of these are
genuine AGN.  However, various lines of evidence to be discussed below
suggest that a significant fraction of these sources do indeed harbour
a LLAGN. We will return to this issue in Section~5.
The detection rate is 25\% for the  MERLIN-detected LINERs.  This
result is  consistent with  the  42\% detection  rate of LINERs  in the
VLA/N00 and VLA/N02 (see also VLA/N05)  distance-limited 
sample, if we consider that
64\% of our MERLIN LINER targets are farther than 20~Mpc.

We  can also  compare the  MERLIN radio detections  for the  subset of
sources (LINERs  and composite galaxies) overlapping with the  galaxy samples 
VLA/N00, VLA/N02 (see also VLA/N05), F00 (see also VLBA/05), HU01, and Filho  \etal  (2004). 
NGC\,2678, NGC\,2841,  NGC\,3169,  NGC\,3190, NGC\,3226, NGC\,3414, NGC\,3607,
NGC\,3627,  NGC\,3628,  NGC\,3718, NGC\,3780, NGC\,3945, NGC\,4143,  NGC\,4203,  NGC\,4293,
NGC\,4450,  NGC\,4548, NGC\,4589, NGC\,4636,  NGC\,4736, NGC\,4772, NGC\,5233, NGC\,5353,
NGC\,5363, NGC\,5377, NGC\,5813 and NGC\,5838,  have been
detected  with the  VLA and/or  VLBA (VLA/N00; VLA/N02; VLA/N05; F00; VLBA/N05). NGC\,777 (HU01) and  
NGC\,5838 (Filho
\etal 2004)  have been detected with  the VLA.  Several sources
(NGC\,3190, NGC\,3607, NGC\,3627, NGC\,3780, NGC\,4548, NGC\,4736 and NGC\,4772) 
went undetected with MERLIN due to
mispointing relative  to the documented radio core.  One (NGC\,4589) perhaps two 
(NGC\,5322; see also Feretti et al. 1984) sources
suffer from severe phase erros. As  for NGC\,2841,  NGC\,3628, NGC\,3945,
NGC\,4636, NGC\,5377, NGC\,5813 (VLA/N002; VLA/N02; VLA/N05), 
NGC\,777 (HU01) and
NGC\,5838  (Filho \etal  2004; VLA/N05) which  have peak  flux  
densities below
2~mJy,  it  is likely  our MERLIN observations  either
resolved out the emission or simply do not have enough signal-to-noise
ratio.  We  note that  the Lovell Telescope  (Jodrell Bank),  the most
sensitive  telescope  of the  MERLIN  array,  was  offline during  our
observations. Of the remaining
sources we  have detected NGC\,2768, NGC\,3169,  NGC\,3226, NGC\,3718, NGC\,4143,
NGC\,4203,  NGC\,4293, NGC\,4450, NGC\,5353, and NGC\,5363 (Table~2 and 3).   



\setcounter{table}{4}

\begin{table*}[!t]

\footnotesize

\begin{center}

\begin{minipage}{100mm}

\caption{Radio data used in the construction of the LTS source RLF.
Col. 1: Spectral class or galaxy.
Col. 2: Number of relevant sources surveyed. In parentheses are the number of 
radio-detected sources. Please note that some sources may be in multiple radio surveys.
Col. 3: Observing frequency.
Col. 4: Resolution.
Col. 5: Instrument.
Col. 6: Flux limit (5$\sigma$).
Col. 7: Reference.}

\begin{tabular}{l c c c c c c}

\hline
\hline

Class/ & & $\nu$ & Res. &   & F$_{\rm limit}$ &  \\

Galaxy & $N$     & (GHz) & (mas)  & Instr.  &  (mJy)  & Ref.      \\

 (1)   & (2)             & (3)   & (4)      & (5)        & (6) & (7) \\

\hline  

Seyfert & 52 (37) & 5 & 1000 & VLA & 0.2 & 1 \\

\hline

LINER & 45 (19) & 15 & 150 & VLA & 1.0 &  2,3 \\

      & 88 (37) & 15 & 150 & VLA & 1.0 & 4 \\


      & 52 (14)  & 5 & 100 & MERLIN & 0.5 & 5 \\






\hline

Composite & 39 (7) &  15 & 150 & VLA & 1.1 &  2,3\\

	  & 62 (9) &  15 & 150 & VLA & 1.0 & 4 \\

          & 16 (1)  & 5   & 100 & MERLIN & 0.5 & 5 \\

          & 35 (26) & 8.4 & 2500 & VLA & 0.3 & 6,7 \\


NGC\,660  & \ldots & 8.4 & 200 & VLA & 0.03 & 8 \\

NGC\,7331 &  \ldots  & 5    & 2000    & VLA   &  0.1   & 9 \\

\hline

\end{tabular}

\smallskip

\scriptsize

{\sc References} -- (1) Ho \& Ulvestad 2001; (2) Nagar \etal 2002; (3)
Nagar \etal  2000; (4) Nagar, Falcke \& Wilson 2005 and references therein;
(5) This paper; (6) Filho, Barthel \& Ho 2002a; (7) Filho, Barthel \& Ho 2000;
(8) data kindly provided by J. Ulvestad; (9) Cowan, Romanishin \& Branch 1994.  
\end{minipage}

\end{center}

\end{table*}

\normalsize


\section{Local Radio Luminosity Function}

In order to construct a representative AGN radio luminosity function (RLF) of the local
Universe, we have used the emission-line (excluding H{\sc ii}) sources in the Palomar 
sample. All of these 
sources  have  now been  observed  at 2\arcsecpoint5~resolution  or
better, when  we include the  present MERLIN observations. 
 We shall refer to the 196 LINERs,
composite  sources  and Seyferts which  satisfy both  the  magnitude 
(B$_{\rm T}<$12.5 mag) and
declination criteria ($\delta>$0\deg) of the Palomar  survey as the `LTS  sources' for
brevity (LTS meaning LINER-Transition-Seyfert); these  sources constitute the present sample.

Ideally,  we   would  like  to   have  a  homogeneous  set   of  radio
observations. But lacking such a  survey, we have assembled in Table~5
a list of radio measurements used to derive the RLF for the LTS sample
sources. Individual  objects taken from  other radio surveys  are also
listed.   Radio  observations   at  different  frequencies  have  been
converted to 5\,GHz  assuming a spectral  index of 0.7
and  also  corrected for  different  cosmologies  if necessary.   When
multiple observations of  the same galaxy were available,  by order of
preference we  choose HU01  for Seyferts,  the radio
observations in  Filho, Barthel  \& Ho (2000,  2002a) and 
VLA/N00, VLA/N02 (see also VLA/N05) for composite galaxies, VLA/N00, VLA/N02
(see also VLA/N05) 
and present MERLIN observations for LINERs.

We  caution that  the  radio-detection rate  will  depend strongly  on
observing   frequency,   resolution   and  sensitivity.    At   higher
frequencies,  many  sources  may  escape  detection  because  spectral
indices are  not always flat and instrumental  sensitivities will also
be lower. Furthermore, the source may suffer resolution effects.  With
this  in mind, the VLA/N00, VLA/N02 samples (see also VLA/N05),  
Filho, Barthel  \& Ho (2000,  2002a)  
and the  present MERLIN observations provide only lower limits  to the 
radio-detection rate in
LTS sources.

Galaxies are considered detected if  their radio flux density is above
5$\sigma$,  where $\sigma$ is  the typical  noise associated  with the
respective survey  (Table~5).  82\%, 43\%,  and 49\% of  the Seyferts,
LINERs,  and  composite  galaxies   were  detected  in  the  radio  at
2\arcsecpoint5~resolution or  less. This is equivalent to  an
overall  radio-detection
rate of 54\% in all LTS  sources or 22\% of all bright nearby galaxies
(Palomar sources). For  Type~1 and 2 LTS sources, the  radio detections 
are 89\%
 and 46\%,  respectively.  Of the Type~1  radio-detected sources 
(which
must be  genuine AGN),  59\% are classified  as Seyferts and  35\% are
LINERs.  Both  broad-lined  composite galaxies (Type  1.9 NGC\,1161 and NGC\,2985)
were  detected.  100\% of Type 1 Seyferts and 67\% of Type 1 LINERs were detected.
46\%,  68\%,  and 94\%  of  the Type~2  radio-detected
sources are Seyferts, LINERs, and composite galaxies, respectively. 

We can also estimate the  fraction of these radio-detected LTS sources
that  are likely to  be genuine  AGN.  In  the absence  of unambiguous
optical  spectral signatures  of  AGN activity  (e.g., broad  emission
lines),  the radio  regime provides  an alternative  and complementary
diagnostic.  Although not a necessary condition for the presence of an
AGN, a compact, flat-spectrum  radio core is indicative of synchrotron
self-absorption,  which  is associated  with  jet  emission from  AGN.
However,  there  are  several  caveats.  First,  ground-based  optical
spectra  and  VLBI-resolution   radio  images  typically  sample  very
different spatial, and presumably temporal, scales.  Thus, optical and
radio signatures of AGN activity need not occur concurrently.  Second,
because  the mechanism  of  jet formation  is  still uncertain,  radio
emission  cannot  be  regarded  as  an inevitable  by-product  of  AGN
activity.  Finally, empirical evidence  suggests that there is a radio
flux  density  threshold  of  2~mJy  below which  the  sources  become
difficult to  detect at milliarcsecond-scale  resolution using current
facilities  (F00; VLBA/N05; Filho \etal
2004).  Known sources  with  submilliJansky radio  
cores and/or  hard
X-ray detections (e.g., NGC\,660  and NGC\,7331; see Filho, Barthel \&
Ho 2002a;  Filho \etal 2004), which  are likely to be  genuine AGN, or
sources  that  are  highly  radio  variable can  be  missed  by  these
relatively shallow,  milliarcsecond-scale resolution observations.  We
therefore  caution that  our estimate  based on  radio detection  is a
lower limit to the true AGN  fraction in LTS sources.  For the sake of
homogeneity,   we   will  restrict   ourselves   to  observations   of
$\ltaprx$1\arcsec~resolution.  With  the above  caveats  in mind,  we
conclude that  at least  80\% of the  Seyferts (HU01),
40\%  of the  LINERs (the  present MERLIN  observations; VLA/N00; VLA/N02; VLA/N05) and 
20\% of  the composite sources (the
present MERLIN observations; VLA/N00; VLA/N02; VLA/N05)
are likely AGN.  Based on the presence of compact radio emission, 
the total fraction of LTS  sources and bright nearby
galaxies   harbouring  AGN   is  therefore   $\sim$50\%   and  $\sim$20\%,
respectively.

According to classical AGN unification schemes (Antonucci 1993), Type 2
objects are simply  Type 1 AGN seen edge-on, whereby the molecular
torus blocks the direct view of the broad-line region (BLR). However, in
the case of LLAGN it is not entirely clear that unification schemes are
readily applicable. There is growing evidence that not all low luminosity
Seyferts, LINERs and composite sources possess a BLR and when they do
the BLR is weak  (Barth, Filippenko \& Moran 1999); 
there are only twelve sources in the Palomar sample (Ho, Filippenko, \& Sargent 1997a)
classified as Type 1.0-1.8 and all are Seyferts. We can then conservatively 
argue that all Type 1 sources, 
which exhibit broad-line emission in their spectra, are genuine AGN. Therefore, 
based solely on the presence of broad-line emission, we can 
estimate that at least 20\% of the LTS sources and 10\% of all bright nearby galaxies
harbour an AGN.

Because the  sample sources  are nearby (median  D = 17~Mpc),  we have
considered  a  flat, Euclidean  Universe  with  q$_0$=0.5 and  H$_{\rm
0}$=75 km  s$^{-1}$ Mpc$^{-1}$  for the subsequent  calculations.  The
V/V$_{\rm  max}$  method  (Schmidt  1968)  was  applied  in  order  to
construct  the RLF  at 5\,GHz.   The main  constraints arise  from the
magnitude limit  of the  Palomar survey which  is taken to  be B$_{\rm
limit}$=12.5~mag (Ho,  Filippenko \& Sargent 1995) and  the radio flux
limit of the survey from where  the radio luminosity of the LTS source
was  obtained  (Table~5).    Moreover,  only  galaxies  with  positive
declination were observed, which  restricts the survey area covered to
2$\pi$.  The calculation of the  RLF is then performed over equal bins
in log  of radio power  (0.4~dex). In each  bin centered on  the radio
luminosity log  L$^*$, the space density (or  differential LF) follows
from the expression:

\begin{center}

$\Phi$(log L$^*$)    =    $\frac{4\pi}{\Omega}   \,    {\Sigma}_{\rm i=1}^{\rm n(log L^*)}
\frac{1}{\rm V_{max (i)}}$,

\end{center}

\noindent where n(log L$^*$) is the number of galaxies in the bin with
luminosities    between    log    L$^*-$0.2   and    log    L$^*+$0.2,
$\frac{4\pi}{\Omega}$  is  the fraction  of  the  sky  covered by  the
optical survey, and V$_{\rm max}$  is the maximum volume of the sphere
in  which the source  could have  been detected  given {\it  both} the
magnitude {\it  and} radio flux limits  ($\equiv\frac{4\pi}{3} \, {\rm
d_{max (i)}}^3$),  with the  maximum volume being  the smaller  of the
two.  Statistical  errors associated  with  the  space densities  were
assigned assuming Poisson statistics.  The total number of objects per
unit volume  brighter than log L$^{*}$  ($\Psi$(log L$^*$); cumulative
version  of  the  RLF) is  obtained  by  summing  over all  the  space
densities in bins with log L$<$log L$^*$:

\begin{center}
 $\Psi$(log L$^*$) =  ${\Sigma}_{\rm i=1}^{\rm n(log L<log L^*)}
\Phi(\rm log L^*)$.
\end{center}

\section {Discussion}

Inspection of Fig.~6a shows  
that space densities of LTS sources continue to  rise with decreasing radio 
power,  with some evidence of flattening 
below  10$^{20}$  W  Hz$^{-1}$, partly due to incompleteness of the 
radio survey (see also discussion in Nagar, Falcke \& Wilson 2005). 
At all radio powers the space densities are clearly dominated
by the LINER galaxies, in particular at the high radio power end
($>$ 10$^{22}$ W  Hz$^{-1}$). Both  Seyfert  and  LINER  galaxies
contribute to  the steady rise  in space densities.  On the other  hand, the
space  densities of composite  sources appear  relatively flat  in the
range 19$<$log  P$_{\rm 5\,GHz}<$21 W  Hz$^{-1}$. 

Table~6 and Fig.~6 contain the derived RLF for the radio-detected 
LTS sample sources, together and separately for for Seyferts, LINERs,
composite sources and then also for Type~1 and Type~2 objects
(with and without broad lines, respectively). It is worth mentioning
that, particularly on the low-power end, the RLF is irregular;
this is the most likely due to density inhomogeneities in our local
volume.

At all radio powers, the  space  densities in  Fig.~6b  are  dominated by  
the  more
numerous Type~2 sources, although 56\%  of the Type~2 sources were not
detected on scales  $<$2\arcsecpoint5 They do, however, span  the same radio
power range as Type~1 sources and appear to turn over at a power
of 10$^{21}$  W Hz$^{-1}$. Type~2 sources roughly  mimic the composite
source and Seyfert behaviour in  its flattening below log~P$_{\rm 5\,GHz} 
= $21 W Hz$^{-1}$.

It is possible that Type~2 or (equivalently composite sources and some
LINERs) are a  mixed case, whereby only a small  fraction of these are
genuine  AGN.  However, various  lines of  evidence, as  stated above,
suggest that  we are  underestimating the number  of LLAGN in  the LTS
sample.  Long  integration X-ray  and radio observations  should prove
useful  in this  regard  to  provide a  complete  survey with  uniform
sensitivity and  resolution.  If there is a  significant population of
submillijansky  LLAGN   that  we   cannot  detect  with   the  present
observations,  then by  including them  in the  RLF, we  should expect
higher space densities, in particular at the low-luminosity end of the
RLF.

Furthermore, because we have  compiled radio measurements from surveys
with different  resolutions, is  it likely that  in some cases  we are
overestimating radio flux densities.  Many of these sources
are  known to  suffer from  resolution effects  (Filho, Barthel  \& Ho
2000, 2002a; VLA/N05; VLBA/N05; Filho \etal 2004); the radio power
of the  underlying AGN may be an  order of magnitude or  so lower than
given by  the arcsec-scale observations.  The overall  effect would be
to shift the RLF to lower radio luminosities.

It is interesting to compare our results with available published surveys.
In VLA/N05 the authors have derived a RLF for a distance-limited sample
of Palomar sources. Comparison between our analysis and theirs shows that the 
results are in rough agreement, within
the errors. In particular, the larger sample of the LTS sources presented
in this study (our 106 versus their 68 radio-detected sources) allows us to 
confirm the low-power turnover seen in the RLF.



\setcounter{table}{5}

\begin{table*}[!t]

\scriptsize

\begin{center}

\begin{minipage}{160mm}

\caption{The local radio luminosity function. 
Col. 1: Binned radio power.
Col. 2,5,8: Number of galaxies in the radio magnitude bin.
Col. 3,6,9: Space density of galaxies and error, calculated assuming Poisson statistics.
Col. 4,7,10: Cumulative number of galaxies.}

\begin{tabular}{c | c c c | c c c | c c c }

\hline
\hline

P   &  & log$\Phi$ & log$\Psi$ &  & log$\Phi$ & log$\Psi$ &  & log $\Phi$ & log$\Psi$   \\

(W Hz$^{-1}$) & $N$ &  (mag$^{-1}$ Mpc$^{-3}$) & (Mpc$^{-3}$) & $N$ &  (mag$^{-1}$ Mpc$^{-3}$) & (Mpc$^{-3}$) & $N$ & (mag$^{-1}$ Mpc$^{-3}$) & (Mpc$^{-3}$) \\

(1)  & (2) & (3) & (4) & (5) & (6) & (7) & (8) & (9) & (10) \\

\hline  

& \multicolumn{3}{l}{Seyferts} & \multicolumn{3}{l}{LINERs} & \multicolumn{3}{l}{Composite}   \\

\hline

$18.2$ & $1$ & $-2.50_{-1.00}^{+0.30}$  & $-2.50$ & \ldots & \ldots & \ldots & \ldots &  \ldots & \ldots  \\

$18.6$  & $2$ & $-3.17_{-0.53}^{+0.23}$ & $-2.41$ & \ldots & \ldots & \ldots &   $2$ & $-2.70_{-0.53}^{+0.23}$ & $-2.70$ \\

$19.0$ & $4$ & $-3.18_{-0.30}^{+0.18}$ & $-2.35$ &  $1$ & $-2.15_{-1.00}^{+0.30}$ & $-2.15$ & $1$ & $-3.58_{-1.00}^{+0.30}$ & $-2.65$  \\

$19.4$ & $6$ & $-2.98_{-0.23}^{+0.15}$ & $-2.25$ & \ldots & $-2.15_{-1.00}^{+0.30}$ & $-2.12$ & $1$ & $-3.43_{-1.00}^{+0.30}$ & $-2.58$  \\

$19.8$ & $4$ & $-3.61_{-0.30}^{+0.18}$ & $-2.24$ & $3$ & $-3.04_{-0.38}^{+0.20}$ & $-2.10$ & $7$ & $-3.26_{-0.21}^{+0.14}$ & $-2.50$  \\

$20.2$ & $4$ & $-3.59_{-0.30}^{+0.18}$ & $-2.22$ & $9$ & $-2.89_{-0.18}^{+0.12}$ & $-2.03$ & $6$ & $-3.63_{-0.23}^{+0.15}$ & $-2.47$ \\

$20.6$ & $3$ & $-3.93_{-0.37}^{+0.20}$ & $-2.21$ & $5$ & $-3.49_{-0.26}^{+0.16}$ & $-2.02$ & $8$ & $-3.40_{-0.19}^{+0.13}$ & $-2.42$  \\

$21.0$ & $5$ & $-4.04_{-0.26}^{+0.16}$ & $-2.20$ & $6$ & $-3.56_{-0.23}^{+0.15}$ & $-2.01$ & $3$ & $-3.82_{-0.37}^{+0.20}$ & $-2.40$  \\

$21.4$ & $4$ & $-4.14_{-0.30}^{+0.18}$ & $-2.20$ & $3$ & $-3.68_{-0.37}^{+0.20}$ & $-2.00$ &$4$  & $-3.31_{-0.30}^{+0.18}$ & $-2.35$  \\

$21.8$ & $2$ & $-4.14_{-0.53}^{+0.23}$ & $-2.19$ & $5$ & $-4.09_{-0.28}^{+0.16}$ & $-2.00$ & \ldots & \ldots & \ldots  \\

$22.2$ & \ldots & \ldots & $-2.19$ & $2$ & $-4.77_{-0.53}^{+0.23}$ & $-2.00$ & \ldots & \ldots &  \ldots   \\

$22.6$ & $1$ & $-4.78_{-1.00}^{+0.30}$ & $-2.19$ & \ldots  &  \ldots & $-2.00$ & \ldots & \ldots & \ldots  \\

$23.0$ & \ldots & \ldots &  $-2.19$ & $1$ & $-5.21_{-1.00}^{+0.30}$ & $-2.00$ & \ldots & \ldots & \ldots  \\

$23.4$ & \ldots & \ldots &  $-2.19$ & $2$ & $5.14_{-0.53}^{+0.23}$ & $-2.00$ & \ldots & \ldots & \ldots  \\

$23.8$ & \ldots & \ldots &  $-2.19$ & \ldots    & \ldots & \ldots & \ldots & \ldots & \ldots  \\

$24.2$ & \ldots & \ldots &  $-2.19$ & \ldots    & \ldots & \ldots & \ldots & \ldots & \ldots  \\

$24.6$ & \ldots & \ldots &  $-2.19$ & \ldots    & \ldots & \ldots & \ldots & \ldots & \ldots  \\

$25.0$ & 1 & $-5.38_{-0.30}^{+1.00}$ & $-2.19$ & \ldots    & \ldots & \ldots & \ldots & \ldots & \ldots  \\

Total & $37$ & \ldots & \ldots & $32$ & \ldots & \ldots & $37$ & \ldots & \ldots \\

\hline

& \multicolumn{3}{l}{Type~1} & \multicolumn{3}{l}{Type~2} & \multicolumn{3}{l}{All}   \\

\hline

$18.2$ & $1$ & $-2.50_{-1.00}^{+0.30}$ & $-2.50$ & \ldots & \ldots & \ldots & $1$ & $-2.50_{-1.00}^{+0.30}$ & $-2.50$  \\

$18.6$ & $1$ & $-3.41_{-1.00}^{+0.30}$ & $-2.45$ & $3$ & $-2.64_{-0.37}^{+0.20}$ & $-2.64$ & $4$ & $-2.57_{-0.30}^{+0.18}$ & $-2.23$  \\

$19.0$ & $1$ & $-3.98_{-1.00}^{+0.30}$ & $-2.44$ & $5$ & $-2.10_{-0.26}^{+0.16}$ & $-1.99$ & $6$ & $-2.10_{-0.23}^{+0.15}$ & $-1.86$  \\

$19.4$ & $3$ & $-3.01_{-0.37}^{+0.20}$ & $-2.33$ & $4$ & $-3.34_{-0.30}^{+0.18}$ & $-1.97$ & $7$ & $-2.86_{-0.21}^{+0.14}$ & $-1.82$ \\

$19.8$ & $2$ & $-3.69_{-0.53}^{+0.23}$ & $-2.31$ & $12$ & $-2.82_{-0.15}^{+0.11}$ & $-1.92$ & $14$ & $-2.77_{-0.14}^{+0.10}$ & $-1.77$  \\

$20.2$ & $8$ & $-3.00_{-0.19}^{+0.13}$ & $-2.23$ & $11$ & $-3.11_{-0.16}^{+0.11}$ & $-1.89$ & $19$ & $-2.75_{-0.11}^{+0.09}$ & $-1.73$   \\

$20.6$ & $4$ & $-3.48_{-0.30}^{+0.18}$ & $-2.21$ & $12$ & $-3.30_{-0.15}^{+0.11}$ & $-1.87$ & $16$ & $-3.08_{-0.13}^{+0.10}$ & $-1.71$  \\

$21.0$ & $7$ & $-3.52_{-0.21}^{+0.14}$ & $-2.19$ & $7$ & $-3.67_{-0.21}^{+0.14}$ & $-1.87$ & $14$ & $-3.29_{-0.14}^{+0.10}$ & $-1.70$  \\

$21.4$ & $1$ & $-3.73_{-1.00}^{+0.30}$ & $-2.18$ & $10$ & $-3.24_{-0.18}^{+0.12}$ & $-1.85$ & $11$ & $-3.11_{-0.16}^{+0.11}$ & $-1.68$  \\

$21.8$ & $3$ & $-4.12_{-0.37}^{+0.20}$ & $-2.17$ & $4$ & $-4.11_{-0.30}^{+0.18}$ & $-1.84$ & $7$ & $-3.81_{-0.21}^{+0.14}$ & $-1.68$  \\

$22.2$ & \ldots & \ldots  & $-2.17$ & $2$ & $-4.77_{-0.53}^{+0.23}$ & $-1.84$ & $2$ & $-4.77_{-0.53}^{+0.23}$ & $-1.68$ \\

$22.6$ & $1$ & $-4.78_{-1.00}^{+0.30}$ & $-2.17$ & \ldots & \ldots & $-1.84$ & $1$ & $-4.78_{-1.00}^{+0.30}$ & $-1.68$ \\

$23.0$ & \ldots & \ldots & $-2.17$ & $1$ & $-5.21_{-1.00}^{+0.30}$ & $-1.84$ & $1$ & $-5.21_{-1.00}^{+0.30}$ & $-1.68$  \\

$23.4$ & $1$ & $-5.56_{-1.00}^{+0.30}$ & $-2.17$ & $1$ & $-5.35_{-1.00}^{+0.30}$ & $-1.84$ & $2$ & $-5.14_{-0.53}^{+0.23}$ & $-1.68$ \\

$23.8$ & \ldots & \ldots               & $-2.17$ & \ldots & \ldots               & \ldots & \ldots & \ldots &  $-1.68$ \\

$24.2$ & \ldots & \ldots &  $-2.17$ & \ldots    & \ldots & \ldots & \ldots& \ldots &  $-1.68$  \\

$24.6$ & \ldots & \ldots &  $-2.17$ & \ldots    & \ldots & \ldots & \ldots & \ldots &  $-1.68$  \\

$25.0$ & $1$ & $-5.38_{-0.30}^{+1.00}$ & $-2.17$ & \ldots    & \ldots & \ldots & $1$ & $-5.38_{-1.00}^{+0.30}$  & $-1.68$  \\
 
Total & $34$ & \ldots & \ldots & $72$ & \ldots & \ldots & $106$ & \ldots & \ldots \\

\hline

\end{tabular}

\end{minipage}

\end{center}

\end{table*}

\normalsize


We have also derived a RLF for the Markarian (Meurs \& Ulvestad 1984) and CfA Seyferts
(Huchra \& Burg 1992; 2\arcsec~resolution VLA radio data from Kukula \etal 1995), converting 
flux densities to 
5\,GHz assuming  $\alpha = 0.7$
(Fig. 7a). We caution that the Markarian Seyferts have been observed with Westerbork
Radio Synthesis Telescope (WRST)
and nuclear radio flux densities may be over-estimated.
The RLFs are consistent with the Palomar Seyfert and RLF
for powers above 10$^{21}$ W Hz$^{-1}$ (Fig. 7a). The rising of the 
LTS and Palomar Seyfert RLF towards lower powers demonstrates that the LTS sample
contains fainter and more local sources than those in the Markarian and CfA
samples. This result is in agreement with that found in VLA/N05
and in HU01, considering the difference in RLF frequency and binning.

Similarly, we can compare our RLF with the AGN sources  in the 2dF  Galaxy Redshift
Survey  of Sadler  \etal (2002).   Galaxies in  this sample  have been
classified as  AGN according  to their spectral  characteristics; they
show  either  an  absorption-line   spectrum  like  that  of  a  giant
elliptical, an absorption spectrum with weak LINER-type emission lines
or  stellar continuum dominated  by nebular  emission lines  of [O{\sc
ii}] or  [O{\sc iii}],  which are strong  compared to  any Balmer-line
emission. The  sample has been cross-correlated with  the NVSS catalog
(Condon \etal 1998).  We have converted the flux  densities to 5\,GHz
assuming an  $\alpha = 0.7$  and corrected for  different cosmologies.
The resulting RLF is plotted along with the LTS RLF in Fig.~7b.

Because the 2dF/NVSS AGN sample is relatively nearby (median $z$=0.2),
it allows  a direct comparison with  the LTS source  RLF.  We caution,
however, that because this  sample has been cross-correlated with NVSS
data, the radio flux densities may be slightly over-estimated.  Fig.~7b
shows that  there is an overlap in radio luminosities for the LTS and 
2dF/NVSS AGN RLF between the regime log P$_{\rm 5\,GHz}\,\approx$ 21
and 23  W Hz$^{-1}$. In this region of 
overlap both the normalization and slope of the two RLFs
are roughly similar, within the errors. The LTS sources naturally extend 
the 2dF/NVSS AGN RLF to 
lower luminosities.  To emphasize the extreme low powers sampled by
our RLF, we note that the lowest power LTS sources  are only 
$\sim$10 times more powerful than Sgr A* (Melia \& Falcke 2001; 
see also Falcke, K\"ording \& Markoff 2004). The overall shape  and the 
smooth  transition from
the 2dF/NVSS AGN RLF to the  LTS RLF, suggest a luminosity continuation  
between  these two source populations. It is natural to view the LTS sources  
as the low-redshift, low-luminosity counterparts of the AGN as sampled by 
the 2dF/NVSS survey.

\section{Conclusions}

We have  undertaken a  MERLIN survey of  nearby galaxies that  did not
have  available  2\arcsecpoint5~resolution  or  better  radio  observations.
Results reveal  a 21\%  radio-detection rate  among the  sources, with
fifteen radio  detections, three of  which are new AGN  candidates.  A
compilation  of  radio observations  of  all low-luminosity  Seyferts,
LINERs and composite LINER/H{\sc ii} galaxies in the magnitude-limited
Palomar survey reveal  a radio-detection rate of 54\%  (or 22\% of all
bright nearby galaxies), with a more than 50\% detection rate (or 20\%
for all bright nearby  galaxies) of low-luminosity active nuclei.  
The radio detection of the  Seyferts, LINERs  and composite  LINER/H{\sc ii}
sources in the Palomar survey  allow the construction of a local radio
luminosity  function.  Our results show  that  the  Seyferts, LINERs  and
composite LINER/H{\sc ii} sources  form a smooth luminosity transition
from higher  redshift, more  luminous AGN as  sampled by  the 2dF/NVSS
survey.

\section{Acknowledgments}

M.~E.~F.  acknowledges support from the Funda\c c\~ao para a Ci\^encia
e Tecnologia,  Minist\'erio da  Ci\^encia e Ensino  Superior, Portugal
through the  grant PRAXIS XXI/BD/15830/98  and SFRH/BPD/11627/2002. We
are  grateful to Jim Ulvestad, Mike  Garrett, Simon  Garrington, Jim  Condon, Naveen
Reddy,  Marco Spaans and  Filippo  Fraternali  for useful
suggestions.   Thanks also  to  Simon Garrington,  Anita Richards  and
Peter Thomasson for valuable help with the data reduction. The authors also
wish to acknowledge the anonymous referee for his insightful suggestions.
 
We have made extensive use of  the FIRST and NVSS online database, the
NASA/IPAC Extragalactic  Database (NED), which is operated  by the Jet
Propulsion  Laboratory,  California  Institute  of  Technology,  under
contract with the National Aeronautics and Space Administration (NASA)
and the Automatic Plate Measuring (APM) Facility, run by the Institute
of Astronomy in Cambridge.  MERLIN  is a national facility operated by
the University of Manchester on behalf of PPARC.


{}



\begin{figure*}
\leavevmode                
\centerline{               
\epsfxsize=8.3cm
\epsffile{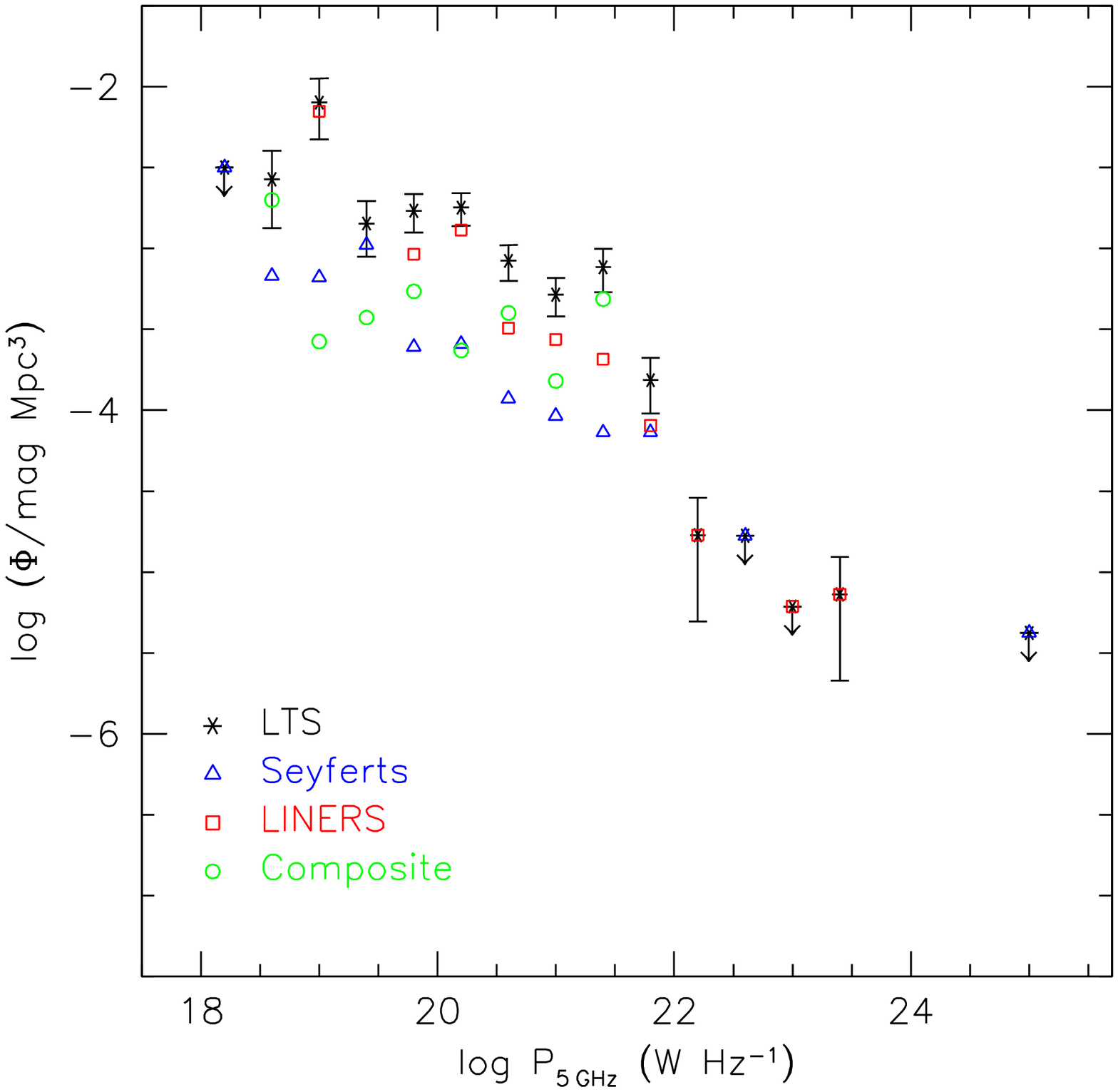}
a) }
\end{figure*}

\begin{figure*}
\leavevmode
\centerline{
\epsfxsize=8.3cm
\epsffile{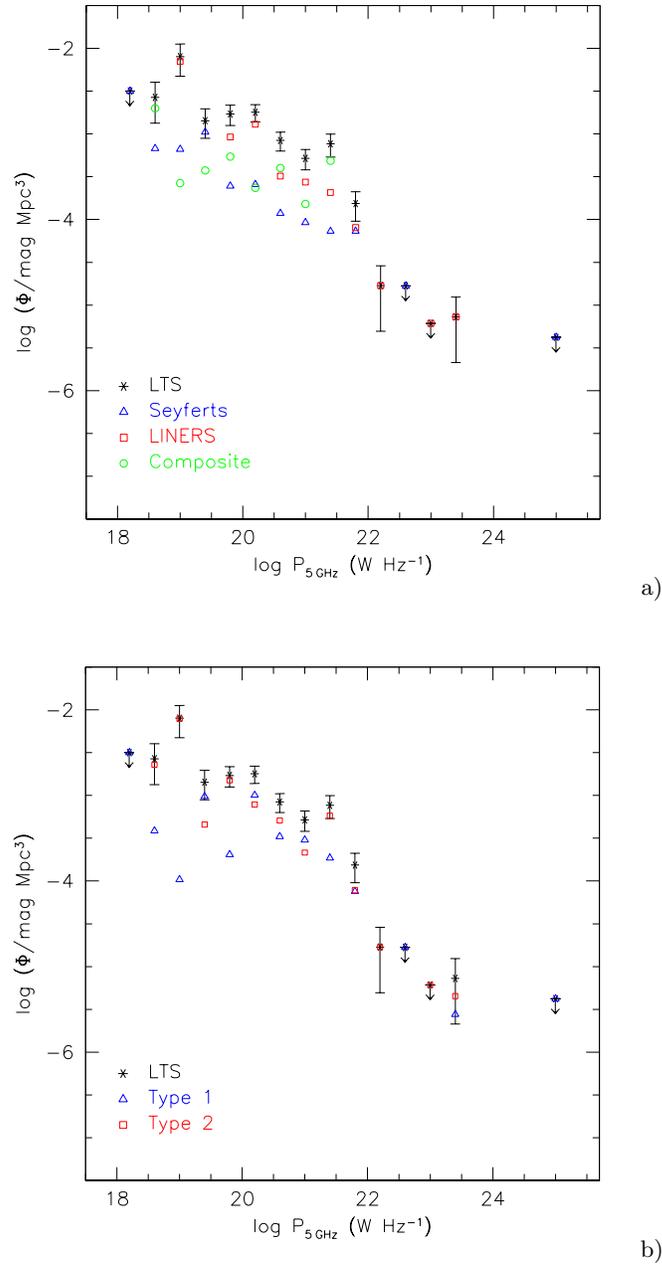}
b)
}

\caption{The  radio luminosity  function for  the radio-detected LTS  sample sources:
  {\bf a)} Seyferts,  LINERs and  composite galaxies  and  {\bf b)}
  Type~1 and Type~2  sources.  Downward arrows are for  bins with only
  one galaxy. Errorbars are assigned assuming Poisson statistics.}
\end{figure*}

\clearpage



\begin{figure*}
\leavevmode
\centerline{
\epsfxsize=8.3cm
\epsffile{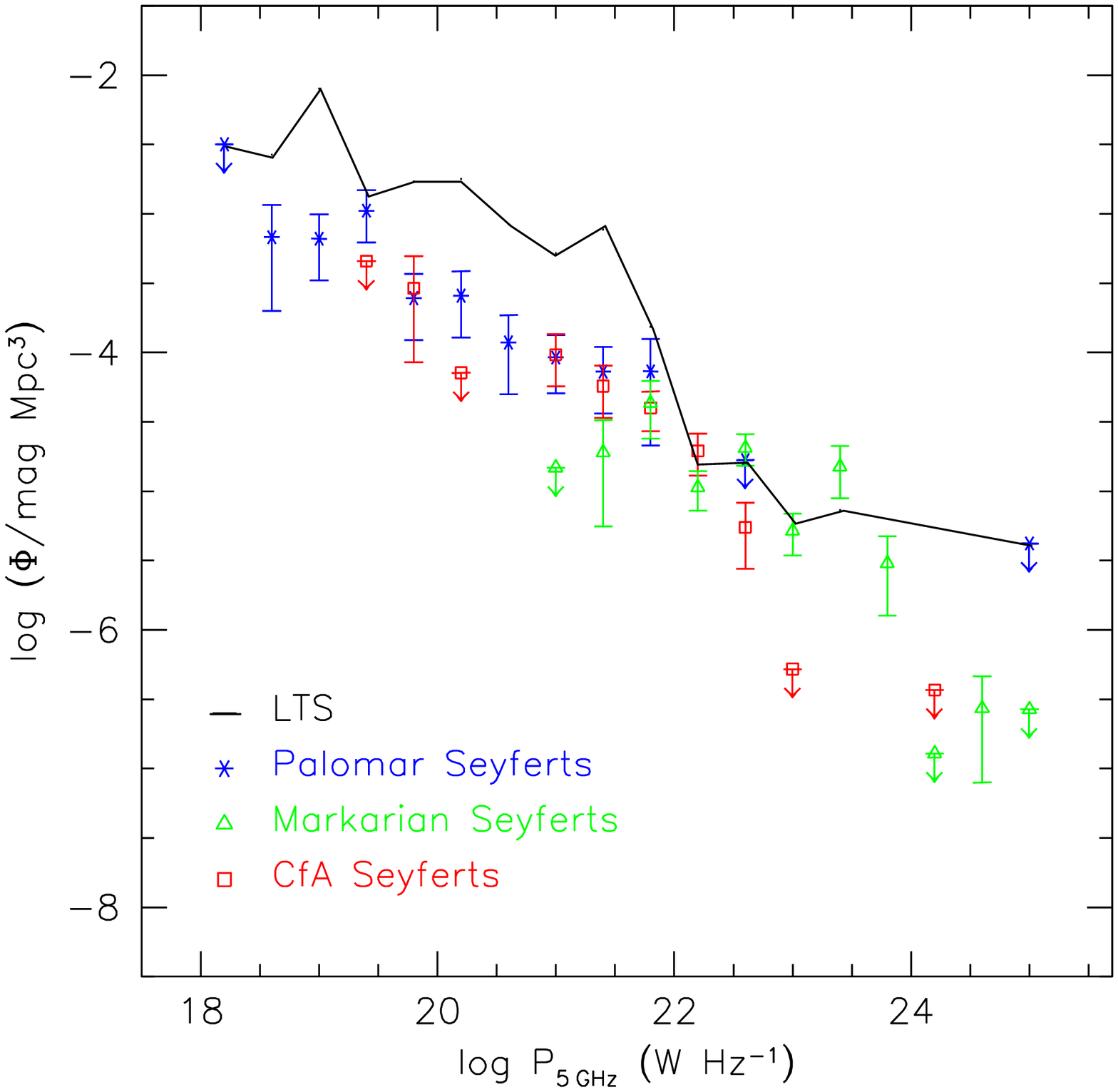}
a) }
\end{figure*}

\begin{figure*}
\leavevmode
\centerline{
\epsfxsize=8.3cm
\epsffile{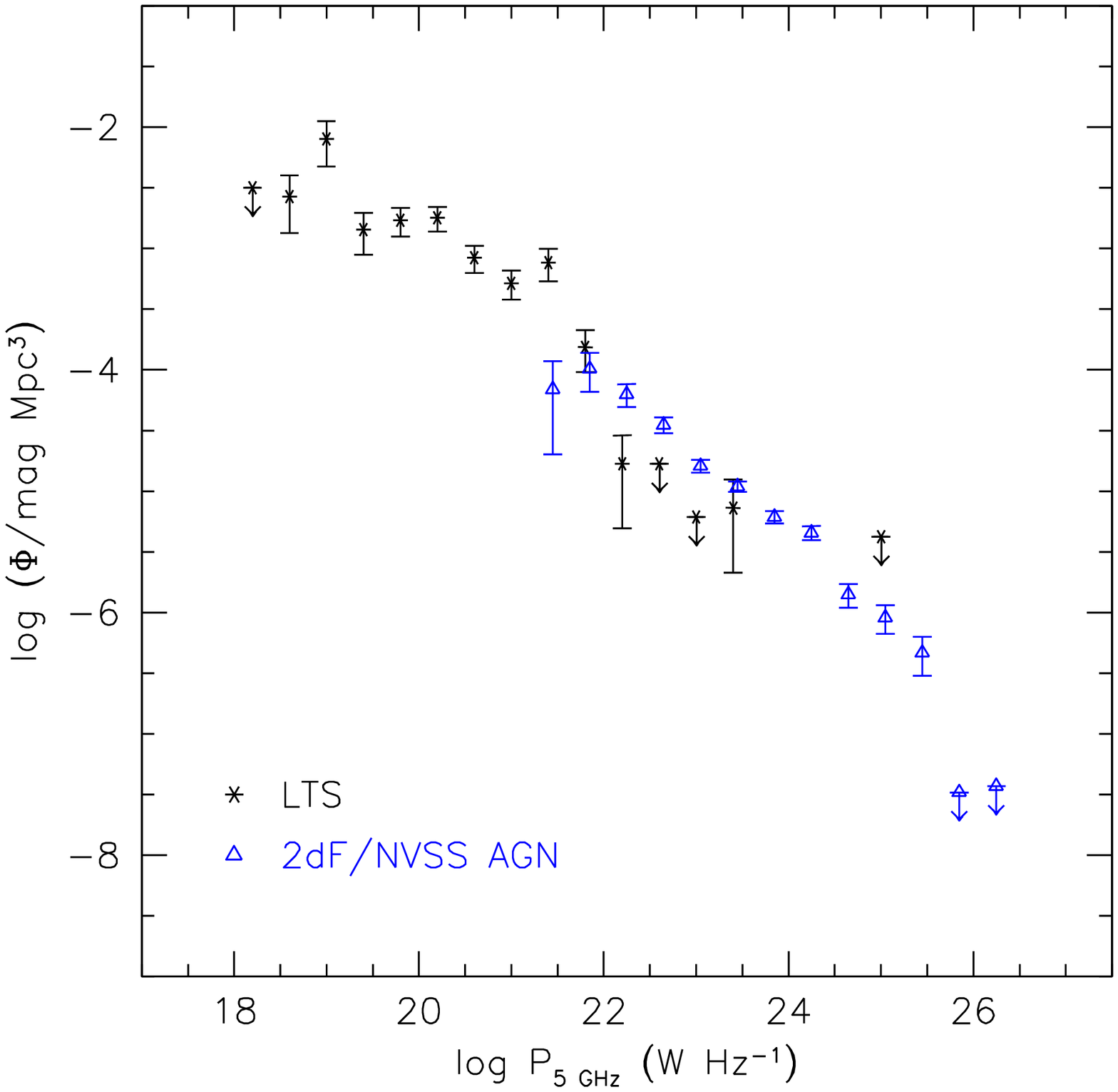}
b)
}

\caption{The radio luminosity function 
for the radio-detected LTS sources compared to: {\bf a)} the Palomar, CfA 
(Huchra \& Burg 1992; radio data from Kukula \etal 1995) and
Markarian Seyferts (Meurs \& Wilson 1984) and {\bf b)}  compared to 2dF/NVSS
AGN (Sadler \etal 2002). Downward arrows are for  bins with only
one galaxy. Errorbars are assigned assuming Poisson statistics.}

\end{figure*}



\begin{thebibliography}{}

\bibitem[]{}
Anderson, J. M., Ulvestad, J. S., \& Ho, L.C. 2004, ApJ, 603, 42

\bibitem[]{}
Antonucci, R. 1993, ARA\&A, 31, 473


\bibitem[]{}
Becker, R.~H., White, R.~L., \& Helfand, D.~J. 1995, ApJ, 450, 559 


\bibitem[]{}
Condon, J.~J. 1992, ARA\&A, 30, 575

\bibitem[]{}
Condon, J.~J., Cotton, W.~D., Greisen, E.~W., Yin, Q.~F., 
Perley, R.~A., Taylor, G.~B., \& Broderick, J.~J. 1998, AJ, 115, 1693

\bibitem[]{}
Cowan, J.~J., Romanishin, W., \& Branch, D. 1994, ApJ, 436, L139

\bibitem[]{}
Falcke, H., K\"ording, E., Markoff, S. 2004, A\&A, 414, 895

\bibitem[]{}
Falcke, H., Nagar, N.~M., Wilson, A.~S., \& Ulvestad, J. 2000, ApJ, 542, 197

\bibitem[]{}
Feretti, L., Giovannini, G., Hummel, E., Kotanyi, G. 1984, A\&A, 137, 362

\bibitem[]{}
Filho, M.~E., Barthel, P.~D., \& Ho, L.~C. 2000, ApJS, 129, 93

\bibitem[]{}
Filho, M.~E., Barthel, P.~D., \& Ho, L.~C. 2002a, ApJS, 142, 223

\bibitem[]{}
Filho, M.~E., Barthel, P.~D., \& Ho, L.~C. 2002b, A\&A, 385, 425

\bibitem[]{}
Filho, M.~E., Fraternali, F., Markoff, S., Nagar, N. M., Barthel, P.~D., Ho, L.~C., \& Yuan, F. 2004, A\&A, 418, 429

\bibitem[]{}
Ho, L.~C. et al. 2001, ApJ, 549, L51

\bibitem[]{}
Ho, L.~C., Filippenko, A.~V., \& Sargent, W.~L.~W. 1995, ApJS, 98, 477

\bibitem[]{}
Ho, L.~C., Filippenko, A.~V., \& Sargent, W.~L.~W. 1997a, ApJS, 112, 315

\bibitem[]{}
Ho, L.~C., Filippenko, A.~V., \& Sargent, W.~L.~W. 1997b, ApJ, 487, 568

\bibitem[]{}
Ho, L.~C.,  \& Ulvestad, J.~S. 2001, ApJS, 133, 77

\bibitem[]{}
Hodge, P. W. \& Kennicutt, R. C. 1983, AJ, 88, 296

\bibitem[]{}
Huchra, J. \& Burg, R. 1992, ApJ, 393, 90

\bibitem[]{}
Kukula, M. J., Pedlar, A., Baum, S. A., \& O'Dea, C. P. 1995, MNRAS, 276, 1262


\bibitem[]{}
Melia, F., Falcke H. 2001, ARA\&A, 39, 309

\bibitem[]{}
Meurs E. J. A. \&  Wilson A. S. 1984, A\&A, 136, 206

\bibitem[]{}
Nagar, N.~M, Falcke, H., Wilson, A.~S., \& Ho, L.~C. 2000, ApJ, 542, 186

\bibitem[]{}
Nagar, N.~M, Falcke, H., \& Wilson, A.~S. 2005, A\&A, 435, 521

\bibitem[]{}
Nagar, N.~M, Falcke, H., Wilson, A.~S., \& Ulvestad, J.~S. 2002, A\&A, 392, 
53


\bibitem[]{}
Sadler, E., et al. 2002, MNRAS, 329, 227

\bibitem[]{}
Schmidt, M. 1968, ApJ, 151, 393

\bibitem[]{}
Terashima, Y., Ho, L.~C., \& Ptak, A.~F. 2000, ApJ, 539, 161 

\bibitem[]{}
Terashima, Y., Ho, L. C., Ptak, A. F., Mushotzky, R. F., Serlemitsos,
P. J., Yaqoob, T., \& Kunieda, H. 2000, ApJ, 533, 729 

\bibitem[]{}
Terashima, Y., \& Wilson, A. S. 2003, ApJ, 583, 145 

\bibitem[]{}
Turner, J. L., \& Ho, P. T. P. 1994, ApJ, 421, 122

\bibitem[]{}
Ulvestad, J.~S., \& Ho, L.~C. 2001a, ApJ, 558, 561

\bibitem[]{}
Ulvestad, J.~S., \& Ho, L.~C. 2001b, ApJ, 562, L133

\bibitem[]{}
Weedman, D. W. 1986, Quasar Astronomy, Cambridge University Press, New YOrk, USA

\end{thebibliography}
\end{document}